\documentclass[aps,pra,twocolumn,superscriptaddress,nofootinbib]{revtex4-2}

\usepackage{amsmath,amssymb,amsfonts,bm}
\usepackage{graphicx}
\usepackage{xcolor}
\usepackage[colorlinks=true,linkcolor=blue,citecolor=blue,urlcolor=blue]{hyperref}

\newcommand{\ii}{\mathrm{i}}
\newcommand{\dd}{\mathrm{d}}
\newcommand{\Tr}{\operatorname{Tr}}
\newcommand{\cL}{\mathcal L}
\newcommand{\cD}{\mathcal D}
\newcommand{\cP}{\mathcal P}
\newcommand{\cQ}{\mathcal Q}
\newcommand{\ket}[1]{|#1\rangle}
\newcommand{\bra}[1]{\langle #1|}
\newcommand{\op}[2]{|#1\rangle\langle #2|}
\newcommand{\bfx}{\mathbf{x}}
\newcommand{\bfy}{\mathbf{y}}
\newcommand{\bfr}{\mathbf{r}}
\newcommand{\bfk}{\mathbf{k}}
\newcommand{\bfp}{\mathbf{p}}
\newcommand{\bfQ}{\mathbf{Q}}
\newcommand{\ehat}{\hat{\mathbf e}}
\newcommand{\Gap}{\Delta_{\mathcal L}}

\begin{document}

\title{Coherence-Mediated Boundary Control of the Liouvillian Gap in Nonreciprocal Open Quantum Systems}

\author{K. L. Li}
\affiliation{School of Physics and Materials Science, Tianjin Normal University, Tianjin 300387, China}

\author{M. H. Yu}
\affiliation{School of Physics and Materials Science, Tianjin Normal University, Tianjin 300387, China}

\author{X. Z. Zhang}
\email{zhangxz@tjnu.edu.cn}
\affiliation{School of Physics and Materials Science, Tianjin Normal University, Tianjin 300387, China}
\affiliation{Interdisciplinary center, Tianjin Normal University, Tianjin
	300387, China}

\begin{abstract}
We study a single-particle Lindblad lattice in which nonreciprocal incoherent hopping is combined with a tunable coherent boundary link.  The link changes the Hamiltonian boundary condition while the incoherent jump rates are kept fixed, so it does not directly alter the diagonal population generator.  Its effect on the Liouvillian gap comes from population-coherence-population feedback: the Hamiltonian couples populations to damped coherences, and eliminating those coherences produces a frequency-dependent self-energy for the slow population branch.  For the periodic reference system, where translation symmetry is imposed on both the Hamiltonian and the dissipator, we derive an exact fixed-momentum reduction, a scalar secular equation for the population-connected branch, and the associated hydrodynamic drift and coherence-corrected diffusion.  For open finite lattices, exact diagonalization and sparse near-zero eigensolvers show that the coherent boundary link can enhance or suppress the gap in one dimension and produce geometry-dependent responses in two dimensions.  We also show that an observed relaxation scale can differ from $1/\Delta_{\mathcal L}$ when the gap family has weak visibility in the chosen perturbation and observable.  The periodic construction gives exact analytical benchmarks, while finite-size simulations in one and two dimensions demonstrate how the same coherence-mediated mechanism controls open-boundary gaps and observable relaxation.
\end{abstract}

\maketitle

\section{Introduction}

The Lindblad master equation provides a standard Markovian description of open quantum systems, in which the spectrum of the Liouvillian superoperator governs relaxation toward the nonequilibrium steady state \cite{Gorini1976,Lindblad1976,BreuerPetruccione2002,1-1PhysRevLett.129.063601,1-2citation,1-3citation,1-4citation}.  If the steady state is unique, the Liouvillian gap is the smallest nonzero decay rate in the real part of the spectrum.  This quantity is central to dissipative state preparation, metastability, and the approach to stationarity.  In nonreciprocal systems, however, the gap cannot be separated from non-Hermitian spectral geometry, boundary sensitivity, and the nonorthogonality of left and right eigenmodes \cite{HatanoNelson1996,YaoWang2018,Kunst2018,Okuma2020,Kawabata2019,Ashida2020,Bergholtz2021,2-1,2-2,2-3}.

Open quantum systems add an important layer to the usual non-Hermitian boundary problem.  Nonreciprocal damping and gain-loss structures can generate chiral damping and dissipative skin accumulation in the dynamics \cite{SongYaoWang2019,Longhi2020,3-1,3-2,3-3}.  In full Lindblad problems, Liouvillian skin effects have been shown to change relaxation without necessarily closing the Liouvillian gap, to appear in exactly solvable quadratic settings, and to produce strong boundary sensitivity of spectra, eigenmodes, and relaxation times \cite{HagaNakagawaHamazakiUeda2021,YangJiangBergholtz2022,ZhouYu2022,FengChen2024,WangLuPengQiWangJie2023,4-1,4-2,4-3,4-4,4-5}.  These developments make the Liouvillian gap a subtle object: it is a spectral decay rate, but observed relaxation can also depend on mode visibility and nonnormal eigenvector geometry \cite{MoriShirai2020,Petermann1979,ChalkerMehlig1998,TrefethenEmbree2005,5-1,5-2,5-3,5-4,5-5}.

The question addressed here is simple but subtle.  Suppose the dissipative jumps are nonreciprocal, while a coherent hopping link at the boundary is tuned from open to periodic.  How can this coherent boundary parameter modify the Liouvillian gap if it is not itself a classical transition rate?  The answer cannot be obtained from the diagonal entries of the density matrix alone.  The diagonal population sector forms a nonreciprocal Markov branch, but the Hamiltonian commutator couples this branch to off-diagonal density-matrix coherences.  These coherences are damped by the dissipator and act as a virtual dissipative channel.  A coherent boundary link changes this channel and feeds back onto the slow population branch through a self-energy.

Let $\cP$ project a density matrix onto its diagonal populations and let $\cQ=1-\cP$ project onto coherences.  If an eigenoperator is written in components as $(p,c)^T$, with $p=\cP\rho$ and $c=\cQ\rho$, the block eigenvalue equation is
\begin{equation}
\begin{pmatrix}
\cL_{pp} & \cL_{pc}\\
\cL_{cp} & \cL_{cc}
\end{pmatrix}
\begin{pmatrix}
p\\
c
\end{pmatrix}
=
z
\begin{pmatrix}
p\\
c
\end{pmatrix}.
\label{eq:block_eigen_main}
\end{equation}
In the block form of the Liouvillian, the direct population block $\cL_{pp}$ is independent of the coherent boundary parameter $\alpha$ for the models considered here.  The $\alpha$ dependence enters the population-coherence and coherence-population blocks.  Whenever $\cL_{cc}-z$ is invertible, the coherence component satisfies
\begin{equation}
c
=
-(\cL_{cc}-z)^{-1}\cL_{cp}p.
\label{eq:coherence_elimination_main}
\end{equation}
Eliminating $c$ therefore gives an effective population problem,
\begin{align}
\cL_{\rm eff}(z,\alpha)
&=
\cL_{pp}-\Sigma(z,\alpha),
\nonumber\\
\Sigma(z,\alpha)
&=
\cL_{pc}(\alpha)
[\cL_{cc}(\alpha)-z]^{-1}
\cL_{cp}(\alpha),
\label{eq:intro_self_energy}
\end{align}
where $z$ is the eigenvalue.  Equation~\eqref{eq:intro_self_energy} is exact whenever the inverse exists.  It makes clear why a boundary condition in the Hamiltonian can control a dissipative gap without directly modifying the jump rates.

\begin{figure}[tbp]
\includegraphics[width=\columnwidth]{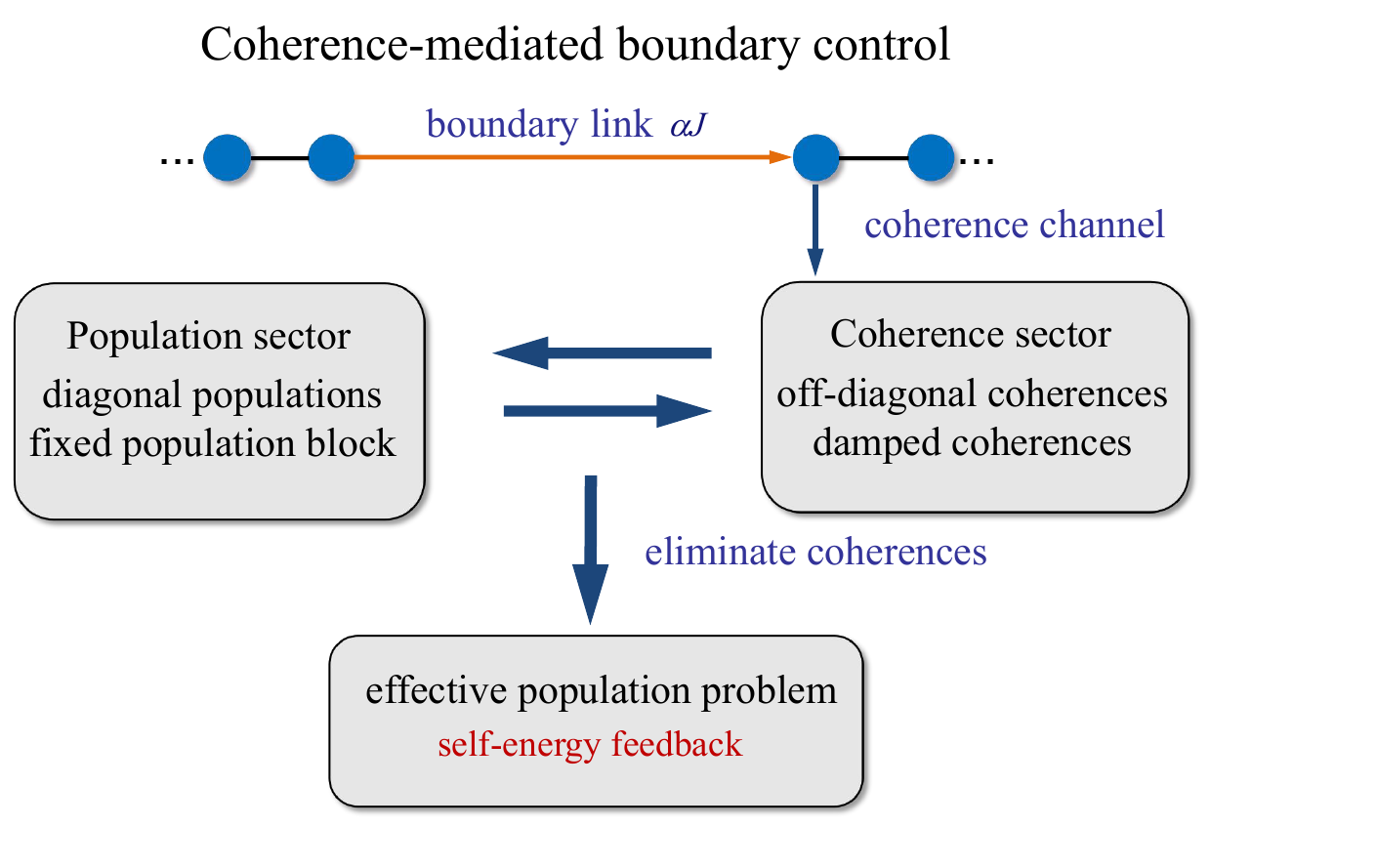}
\caption{Mechanism of coherent boundary control.  The diagonal population sector $\cP\rho$ carries the classical nonreciprocal branch, while the off-diagonal sector $\cQ\rho$ contains damped coherences.  The coherent boundary link $\alpha J$ changes the coherence channel.  Eliminating the coherence sector gives the self-energy in Eq.~\eqref{eq:intro_self_energy}.  Thus $\alpha$ changes the Liouvillian gap through $\Sigma(z,\alpha)$ rather than through a direct change of $\cL_{pp}$.}
\label{fig:mechanism}
\end{figure}

This mechanism is summarized in Fig.~\ref{fig:mechanism}.  There is an additional point that is experimentally relevant.  The Liouvillian gap is a spectral quantity.  The relaxation time extracted from a chosen initial perturbation and a chosen observable also depends on mode visibility.  In a non-normal Liouvillian, the right and left eigenoperators associated with the slow mode may be localized differently.  Their overlap, measured for example by a Petermann-type factor, and their projection onto the physical protocol can make the observed relaxation differ from $1/\Gap$.

In this work we develop a finite-size analytical and numerical framework for this mechanism in one and two dimensions.  The periodic reference system is treated analytically by an exact fixed-momentum reduction.  The open-boundary systems are analyzed numerically with explicit validations against dense diagonalization, trace preservation, and block diagnostics.  Exact results below refer to algebraic block identities, the periodic reference construction, and full finite-size diagonalization.  The boundary-response curves and relaxation fits are finite-size numerical evidence.  We do not infer a thermodynamic-limit scaling law from these finite lattices, and we do not claim a full Liouvillian topological invariant.

The remainder of the paper is organized as follows.  Section~II defines the model, the boundary conditions, and the Liouvillian gap.  Section~III gives the periodic fixed-momentum reduction and the hydrodynamic expansion.  Section~IV analyzes one-dimensional coherent boundary control through the Feshbach self-energy, and Sec.~V extends the same mechanism to two-dimensional geometries.  Section~VI compares the spectral inverse gap with observed relaxation.  Section~VII concludes the paper and outlines possible extensions.  Technical derivations and numerical checks are collected in the Appendices.

\section{Model and Gap}

We consider a single particle on a $d$-dimensional hypercubic lattice.  A site is denoted by $\bfx=(x_1,\ldots,x_d)$, with $1\leq x_\mu\leq L_\mu$, and $\ehat_\mu$ is the unit vector in direction $\mu$.  The density matrix obeys
\begin{align}
\frac{\dd \rho}{\dd t}
&=
\cL\rho
=
-\ii[H(\bm{\alpha}),\rho]
\nonumber\\
&\quad+
\sum_{\bfx,\mu,s=\pm}
\left(
L_{\bfx,\mu,s}\rho L_{\bfx,\mu,s}^\dagger
-\frac{1}{2}\{L_{\bfx,\mu,s}^\dagger L_{\bfx,\mu,s},\rho\}
\right).
\label{eq:lindblad}
\end{align}
Here the jump operators are nearest-neighbor incoherent hoppings,
\begin{align}
L_{\bfx,\mu,+}
&=
\sqrt{\gamma_{\mu,+}}\,
\op{\bfx+\ehat_\mu}{\bfx},
\nonumber\\
L_{\bfx,\mu,-}
&=
\sqrt{\gamma_{\mu,-}}\,
\op{\bfx-\ehat_\mu}{\bfx},
\label{eq:jump_ops}
\end{align}
whenever the corresponding bond exists.  The rates $\gamma_{\mu,+}$ and $\gamma_{\mu,-}$ need not be equal.  Their ratio
\begin{equation}
r_\mu=\frac{\gamma_{\mu,+}}{\gamma_{\mu,-}}
\label{eq:rate_ratio}
\end{equation}
sets the direction and strength of dissipative nonreciprocity along direction $\mu$.

The coherent Hamiltonian is Hermitian.  In one dimension,
\begin{align}
H(\alpha)
&=
J\sum_{x=1}^{N-1}
\left(
\op{x+1}{x}+\op{x}{x+1}
\right)
\nonumber\\
&\quad+
J\left(
\alpha\op{1}{N}+\alpha^*\op{N}{1}
\right).
\label{eq:H_1d_alpha}
\end{align}
The real scans below use $0\leq\alpha\leq1$.  The limits $\alpha=0$ and $\alpha=1$ correspond to an open and a closed coherent Hamiltonian link, respectively.  In these open-boundary scans the incoherent jump network is held fixed: the jump operators in Eq.~\eqref{eq:jump_ops} are defined only on the bonds that exist in the open lattice, and varying $\alpha$ changes only the Hamiltonian boundary link.  Thus $\alpha=1$ does not by itself make the dissipator periodic.  The fully periodic Liouvillian used in Sec.~III is a separate reference problem in which translation symmetry is imposed on both the Hamiltonian and the dissipator.  In two dimensions we use independent coherent boundary parameters $\alpha_x$ and $\alpha_y$ for the $x$ and $y$ directions.  On a rectangle they close the corresponding coherent boundary links when present; on a cylinder only the chosen direction is closed.

The Liouvillian gap is defined as
\begin{equation}
\Gap
=
-\max_{\lambda_n\neq0}\operatorname{Re}\lambda_n,
\label{eq:gap_definition}
\end{equation}
where $\lambda_n$ are the finite-size Liouvillian eigenvalues.  The zero eigenvalue corresponds to the steady state.  In all scans reported here the numerical zero-mode count is one within the stated tolerance.  The gap mode may be real or may belong to a complex-conjugate pair.  In both cases $\Gap$ is extracted from the real part in Eq.~\eqref{eq:gap_definition}.  The vectorization convention and the finite Liouvillian matrix used for this extraction are given in Appendix~\ref{app:vectorization}, and the numerical validation protocol is summarized in Appendix~\ref{app:numerics}.
In the figures, the compact label $\Delta_{\rm L}$ denotes the same Liouvillian gap $\Gap$.

\section{Periodic Reference System}

The periodic system provides a useful exact benchmark because translational symmetry block diagonalizes the Liouvillian.  Define a Liouville-space basis
\begin{equation}
O_{\bfQ,\bfr}
=
\sum_{\bfx}
e^{\ii\bfQ\cdot\bfx}
\op{\bfx+\bfr}{\bfx},
\label{eq:O_Qr}
\end{equation}
where $\bfQ$ is the center-of-mass momentum and $\bfr$ is the relative coordinate between ket and bra.  This relative coordinate should not be confused with the nonreciprocity ratio $r_\mu$ in Eq.~\eqref{eq:rate_ratio}.  The subspace with $\bfr=0$ is the population sector, and the subspace with $\bfr\neq0$ is the coherence sector.  The dissipator acts on the population state as
\begin{equation}
\cD O_{\bfQ,0}
=
a_{\bfQ}O_{\bfQ,0},
\label{eq:D_pop_branch}
\end{equation}
with
\begin{equation}
a_{\bfQ}
=
\sum_{\mu=1}^d
\left[
\gamma_{\mu,+}(e^{-\ii Q_\mu}-1)
+
\gamma_{\mu,-}(e^{\ii Q_\mu}-1)
\right].
\label{eq:a_Q}
\end{equation}
This is the Bloch eigenvalue of a classical nonreciprocal Markov process.  For $\bfr\neq0$, the jump-gain term vanishes and the coherences decay with the total exit rate
\begin{equation}
\Gamma=\sum_{\mu=1}^d(\gamma_{\mu,+}+\gamma_{\mu,-}).
\label{eq:Gamma_total}
\end{equation}
The Hamiltonian then couples $\bfr=0$ to neighboring coherences and, recursively, to longer-range coherences.  This is the microscopic origin of the self-energy in Eq.~\eqref{eq:intro_self_energy}.  The fixed-$\bfQ$ block derivation is given in Appendix~\ref{app:periodic_reduction}, and the hydrodynamic expansion is worked out in Appendix~\ref{app:momentum_hydro}.

For the periodic reference Hamiltonian
\begin{equation}
\varepsilon_{\bfk}
=
2\sum_{\mu=1}^d J_\mu\cos k_\mu,
\label{eq:dispersion}
\end{equation}
the population-connected branch in a fixed-$\bfQ$ sector satisfies
\begin{equation}
1
=
c_{\bfQ}
\frac{1}{V}\sum_{\bfp}
\frac{1}{
z+\Gamma+\ii[\varepsilon_{\bfp+\bfQ}-\varepsilon_{\bfp}]
},
\label{eq:secular}
\end{equation}
where
\begin{equation}
c_{\bfQ}
=
\sum_{\mu=1}^d
\left(
\gamma_{\mu,+}e^{-\ii Q_\mu}
+
\gamma_{\mu,-}e^{\ii Q_\mu}
\right).
\label{eq:c_Q}
\end{equation}
When all $J_\mu=0$, Eq.~\eqref{eq:secular} reduces exactly to $z=a_{\bfQ}$ for this population-connected branch.

For small $\bfQ$, this population-connected branch has the hydrodynamic expansion
\begin{equation}
z_{\bfQ}
=
-\ii\sum_\mu v_\mu Q_\mu
-
\sum_{\mu,\nu}D_{\mu\nu}^{\rm eff}Q_\mu Q_\nu
+O(Q^3),
\label{eq:hydro}
\end{equation}
with drift velocity
\begin{equation}
v_\mu=\gamma_{\mu,+}-\gamma_{\mu,-},
\label{eq:drift}
\end{equation}
and effective diffusion tensor
\begin{equation}
D_{\mu\nu}^{\rm eff}
=
\frac{\gamma_{\mu,+}+\gamma_{\mu,-}}{2}\delta_{\mu\nu}
+
\frac{1}{\Gamma}
\left\langle
\partial_{p_\mu}\varepsilon_{\bfp}\,
\partial_{p_\nu}\varepsilon_{\bfp}
\right\rangle_{\bfp}.
\label{eq:D_eff_general}
\end{equation}
For the nearest-neighbor cosine band this becomes
\begin{equation}
D_{\mu\nu}^{\rm eff}
=
\left[
\frac{\gamma_{\mu,+}+\gamma_{\mu,-}}{2}
+
\frac{2J_\mu^2}{\Gamma}
\right]\delta_{\mu\nu}.
\label{eq:D_eff_cosine}
\end{equation}
The second term is the controlled coherence-mediated correction.  It scales as $J_\mu^2/\Gamma$ because the Hamiltonian creates coherences with amplitude $J_\mu$ and the dissipator damps them on the time scale $\Gamma^{-1}$.

\begin{figure*}[t]
\includegraphics[width=0.95\textwidth]{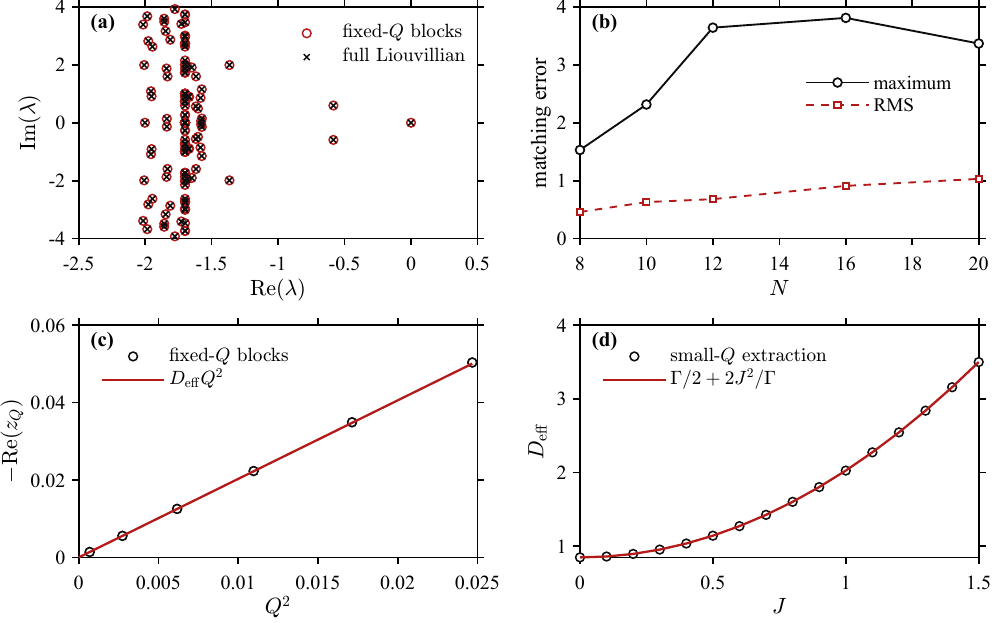}
\caption{Periodic one-dimensional benchmark.  (a) Full Liouvillian spectrum and the union of fixed-$Q$ block spectra for $N=12$, $J=1$, $\gamma_+=1.3$, and $\gamma_-=0.4$.  (b) Maximum and RMS matching errors between the two constructions as a function of $N$, plotted in units of $10^{-14}$.  (c) Small-$Q$ decay rate $-\operatorname{Re}z_Q$ compared with $D_{\rm eff}Q^2$ for $N=240$.  (d) Extracted effective diffusion as a function of $J$, compared with $\Gamma/2+2J^2/\Gamma$ in one dimension, where $\Gamma=\gamma_++\gamma_-$.}
\label{fig:pbc_benchmark}
\end{figure*}

Figure~\ref{fig:pbc_benchmark} verifies the periodic construction in one dimension.  The full $N^2\times N^2$ spectrum and the fixed-$Q$ block spectra agree within double precision for $N=12$, $J=1$, $\gamma_+=1.3$, and $\gamma_-=0.4$: the maximum and RMS spectrum-matching errors are below $4\times10^{-14}$ and $7\times10^{-15}$.  For $J=0$ the extracted branch agrees with Eq.~\eqref{eq:a_Q} to within $10^{-15}$.  The small-$Q$ fit at $N=240$ gives $D_{\rm eff}\simeq2.03$, consistent with Eq.~\eqref{eq:D_eff_cosine}; the fitted drift is $0.900$, consistent with $\gamma_+-\gamma_-=0.9$.  The detailed benchmark numbers are recorded in Appendix~\ref{app:numerics}.

The periodic calculation is not used as a direct formula for the open-chain gap.  Its role is to provide an exact benchmark showing how damped coherences generate a self-energy correction to a population branch.  In the open chain, the coherent boundary link modifies the same coherence-return channel in a finite geometry.

\section{One-Dimensional Boundary Control}

\begin{figure*}[t]
\includegraphics[width=0.95\textwidth]{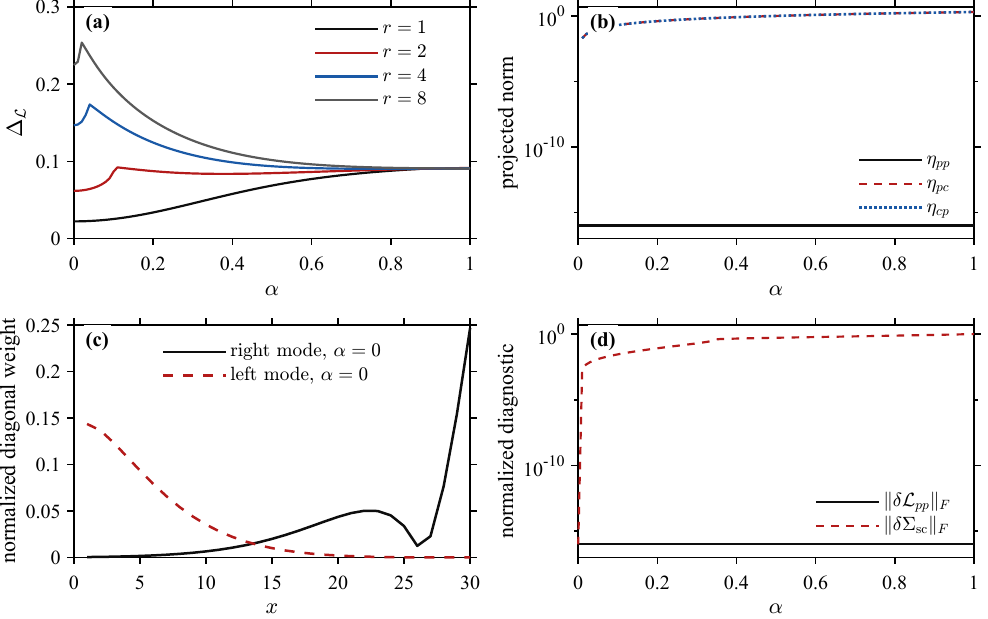}
\caption{One-dimensional coherence-mediated boundary response for $N=30$, $J=1$, and $\Gamma=1.7$.  The rates are $\gamma_+=\Gamma r/(1+r)$ and $\gamma_-=\Gamma/(1+r)$.  (a) Liouvillian gap as a function of the coherent boundary parameter $\alpha$ for several nonreciprocity ratios $r$.  (b) Projected block norms showing that $\cL_{pp}$ is independent of $\alpha$, while population-coherence couplings change with $\alpha$.  (c) Normalized diagonal weights of the gap-mode right and left eigenoperators for $r=4$ at $\alpha=0$.  (d) Feshbach diagnostic for $N=12$ and $r=4$: the plotted self-energy difference is $\delta\Sigma_{\rm sc}(\alpha)=\Sigma[z_{\rm gap}(\alpha),\alpha]-\Sigma[z_{\rm gap}(0),0]$, and the direct population block is fixed.}
\label{fig:boundary_alpha}
\end{figure*}

We now turn to the open chain with the coherent boundary link in Eq.~\eqref{eq:H_1d_alpha}.  The important structural fact is that changing $\alpha$ does not modify the direct population block $\cL_{pp}$.  It modifies $\cL_{pc}$, $\cL_{cp}$, and $\cL_{cc}$, and hence the self-energy
\begin{equation}
\Sigma(z,\alpha)
=
\cL_{pc}(\alpha)
[\cL_{cc}(\alpha)-z]^{-1}
\cL_{cp}(\alpha).
\label{eq:sigma_boundary_section}
\end{equation}
The matrix-unit derivation of this statement is given in Appendix~\ref{app:feshbach}.  The numerical block diagnostic in Fig.~\ref{fig:boundary_alpha}(b) confirms
\begin{equation}
\|\cL_{pp}(\alpha)-\cL_{pp}(0)\|_F=0
\label{eq:Lpp_invariant}
\end{equation}
within numerical precision, whereas the population-coherence norms grow linearly with $\alpha$.  For the representative $N=30$, $\Gamma=1.7$, $J=1$, and $r=4$ calculation, the fitted projected norms satisfy $\eta_{pc}=\eta_{cp}=2\alpha$ and $\eta_{pp}=0$.

The Frobenius norm in Fig.~\ref{fig:boundary_alpha}(d) is a channel diagnostic rather than a scalar predictor of the gap.  It verifies the route by which $\alpha$ enters the effective population problem, while the nonmonotonic gap in Fig.~\ref{fig:boundary_alpha}(a) is controlled by the projection of this self-energy onto the selected slow branch and by possible slow-family selection.

The effect on the gap is not monotonic in the nonreciprocity ratio.  Figure~\ref{fig:boundary_alpha}(a) shows $\Gap(\alpha)$ for $r=1,2,4,8$ at $N=30$, $J=1$, and $\Gamma=1.7$, where $\gamma_+=\Gamma r/(1+r)$ and $\gamma_-=\Gamma/(1+r)$.  For $r=1$, the gap increases from $0.0223$ at $\alpha=0$ to a maximum $0.0906$ near $\alpha=0.89$.  For $r=4$, the open-chain value is already larger, $\Gap(0)=0.147$, and the maximum $0.174$ occurs near $\alpha=0.04$.  For $r=8$, the maximum is $0.254$ near $\alpha=0.02$, while the closed coherent link gives $\Gap(1)=0.0908$.  Table~\ref{tab:boundary_gaps} gives the full finite-size values.

Equivalently, if $z_n(\alpha)$ denote the nonzero finite-size Liouvillian eigenvalues,
\begin{equation}
\Gap(\alpha)
=
\min_{n\neq0}
\left[
-\operatorname{Re}z_n(\alpha)
\right].
\label{eq:gap_lower_envelope}
\end{equation}
The observed gap curve is the lower envelope of the decay rates of all nonzero Liouvillian branches.  A peak can therefore arise either from the curvature of a single selected slow branch or from a change in the branch that controls the lower envelope.  This explains why the nonmonotonicity in Fig.~\ref{fig:boundary_alpha}(a) should not be interpreted as a simple monotonic response of one fixed eigenvalue.

The peak is also not predicted by the periodic hydrodynamic coefficient $D_{\rm eff}$.  It is a finite-size response to a coherent boundary perturbation.  Near the open-chain point one can write
\begin{equation}
\cL(\alpha)
=
\cL_0+\alpha\mathcal V_{\rm bd},
\label{eq:weak_link_main}
\end{equation}
with $\cL_0=\cL(0)$.  For an isolated slow branch $g$ of $\cL_0$, non-Hermitian perturbation theory gives
\begin{equation}
z_g(\alpha)
=
z_g^{(0)}
+
\alpha z_g^{(1)}
+
\alpha^2 z_g^{(2)}
+
O(\alpha^3).
\label{eq:branch_weak_link_main}
\end{equation}
If this same branch remains selected by the lower envelope, the local single-branch stationary point is estimated by
\begin{equation}
\alpha_{\max}^{(g)}
\simeq
-
\frac{\operatorname{Re}z_g^{(1)}}
{2\operatorname{Re}z_g^{(2)}}.
\label{eq:alpha_max_weak_link_main}
\end{equation}
The detailed expressions for $z_g^{(1)}$ and $z_g^{(2)}$ are given in Appendix~\ref{app:feshbach}.  For the $N=30$ data in Fig.~\ref{fig:boundary_alpha}(a), the biorthogonal calculation gives an almost vanishing $z_g^{(1)}$ and a positive quadratic correction to the decay rate, with $-\operatorname{Re}z_g^{(2)}$ increasing from $0.315$ at $r=1$ to $37.5$ at $r=8$.  Thus strong nonreciprocity makes a very small coherent boundary link produce an $O(10^{-2})$ gap enhancement.  The same calculation does not give a reliable closed-form $\alpha_{\max}$ for the observed maxima: the reciprocal maximum near $\alpha\simeq0.89$ is outside the weak-link regime, and for $r=4,8$ the controlled expansion describes the initial rise but not the subsequent finite-size turnover.

Together, these observations give Fig.~\ref{fig:boundary_alpha} a specific microscopic meaning.  The nonreciprocal jumps generate a boundary-sensitive population branch, while the coherent Hamiltonian dresses that branch through damped coherences.  The boundary link can initially increase the gap by changing the virtual return paths in the coherence sector; at larger $\alpha$ the selected slow family can turn over or reorganize.  The response is therefore a spectral consequence of the self-energy, not a direct change of a classical population transition rate.

For an isolated non-Hermitian eigenvalue, the first-order sensitivity to a real boundary parameter is controlled by the biorthogonal boundary matrix element.  Let $R_n$ and $L_n$ denote the corresponding right and left eigenoperators, viewed as Liouville-space vectors with eigenvalue $z_n$.  Then
\begin{equation}
\partial_\alpha z_n
=
\frac{(L_n|\partial_\alpha\cL|R_n)}
{(L_n|R_n)}.
\label{eq:eigenvalue_sensitivity}
\end{equation}
With biorthogonal normalization this reduces to $\partial_\alpha z_n=(L_n|\partial_\alpha\cL|R_n)$.  Since the left and right gap modes can have different spatial weights, the boundary matrix element in Eq.~\eqref{eq:eigenvalue_sensitivity} is mode dependent.  This local sensitivity explains why the initial gap response can depend strongly on the nonreciprocity ratio.  In the present one-dimensional scans, the lower-envelope viewpoint mainly cautions against reading the full gap curve as a monotonic response of one eigenvalue.  For $r=4$ and $r=8$, the branch tracking summarized in Appendix~\ref{app:feshbach} indicates that the same gap family controls the small-$\alpha$ peak region; the peak is therefore a finite-size turnover of this selected family beyond the local weak-link expansion.  In the two-dimensional geometries of Sec.~V, by contrast, the selected slow channel can rearrange along a boundary-link cut.

The left and right eigenoperators further show the non-normal geometry.  For $N=30$, $r=4$, and $\alpha=0$, the gap-mode right diagonal weight has center of mass $24.3$, while the left diagonal weight has center of mass $5.40$.  The Petermann factor is $K\simeq9.81\times10^2$.  At $\alpha=1$, the corresponding centers of mass are $16.2$ and $15.3$, and $K\simeq1.22$.  The boundary link therefore changes both the eigenvalue and the geometry of the slow mode.  The precise definitions of the left-right weights and the Petermann factor are given in Appendix~\ref{app:relaxation}.

\section{Two-Dimensional Geometry}

In two dimensions the coherent boundary response depends on the direction of nonreciprocity and on the geometry of the opened boundaries.  We use a $6\times6$ rectangle with $J_x=J_y=1$ and $\Gamma_x=\Gamma_y=1.7$.  The nonreciprocity ratios are $r_x=\gamma_{x,+}/\gamma_{x,-}$ and $r_y=\gamma_{y,+}/\gamma_{y,-}$.  The coherent boundary parameters $\alpha_x$ and $\alpha_y$ close the $x$ and $y$ Hamiltonian links, respectively, while the dissipative jump network remains open in the corresponding open directions.

\begin{figure*}[t]
\includegraphics[width=0.96\textwidth]{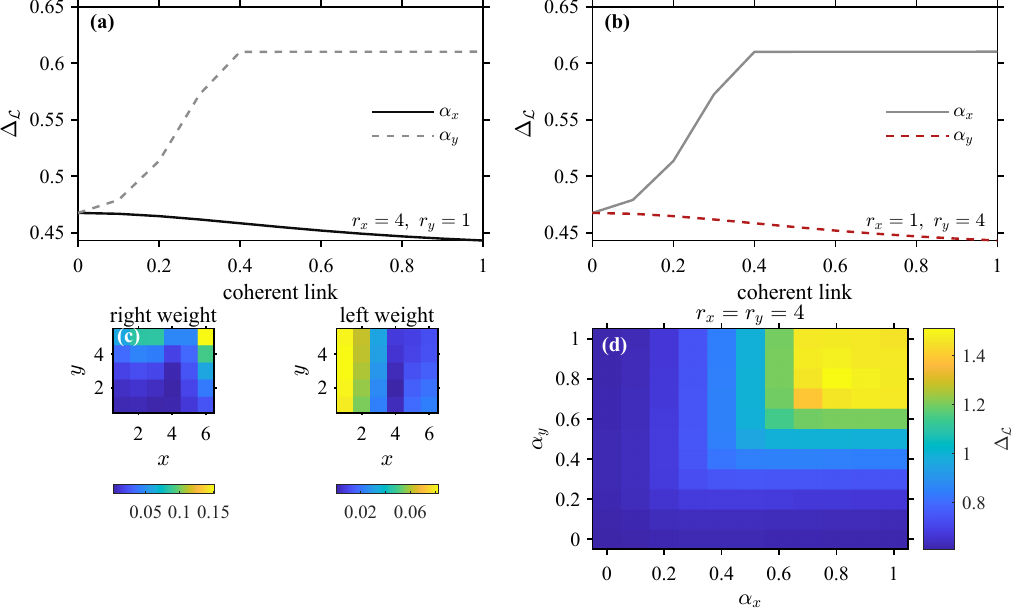}
\caption{Geometry-dependent boundary response in two dimensions.  (a) Open $6\times6$ rectangle with $(r_x,r_y)=(4,1)$, showing the gap along $\alpha_x$ and $\alpha_y$ cuts.  (b) Same geometry with $(r_x,r_y)=(1,4)$.  In (a) and (b), the global gap is selected by the slowest finite-size channel, so the dominant response can be controlled by slow-channel competition rather than by a single microscopic hopping ratio.  (c) Normalized right and left diagonal weights of the gap mode for a $6\times5$ rectangle in the corner-biased case $(r_x,r_y)=(4,4)$ at $\alpha_x=\alpha_y=0$.  (d) Gap map for the open $6\times6$ rectangle with $(r_x,r_y)=(4,4)$, showing the cooperative enhancement when both coherent links are varied.  The point-gap winding used in the analysis is a diagnostic of the classical population branch on a torus, not a proof of a full Liouvillian topological invariant.}
\label{fig:geometry_2d}
\end{figure*}

Figure~\ref{fig:geometry_2d}(a) shows case A, with $(r_x,r_y)=(4,1)$.  In the plotted open rectangle the global gap responds more strongly to the transverse coherent link: $R_x\simeq0.0244$ and $R_y\simeq0.143$, where $R_\mu$ denotes the maximum gap change along the corresponding cut.  The initial slopes are $\chi_x\simeq-0.00778$ and $\chi_y\simeq0.115$.  Rotating the nonreciprocity to case B, $(r_x,r_y)=(1,4)$, rotates this finite-size response, as shown in Fig.~\ref{fig:geometry_2d}(b): $R_x\simeq0.143$ and $R_y\simeq0.0244$.  These data do not contradict skin-type boundary sensitivity.  The transverse response in the open rectangle reflects which finite-size slow family controls the lower envelope of decay rates.

The useful physical interpretation is slow-channel competition.  The Liouvillian gap is a global spectral quantity rather than a local measure of one selected skin mode.  The global finite-size gap is the minimum decay rate among several geometry-dependent channels, not a response coefficient tied to a single microscopic hopping ratio.  Therefore the direction with stronger microscopic nonreciprocity need not be the direction with the strongest response of the global gap.  In case A, the transverse coherent link dominates because it modifies the slow channel that selects the gap in this finite geometry.  The open-rectangle response in Figs.~\ref{fig:geometry_2d}(a) and \ref{fig:geometry_2d}(b) should therefore be viewed as the slow-channel competition case.  Appendix~\ref{app:geometry} shows that the responsive transverse cut is accompanied by a rearrangement of the lowest few decay rates near $\alpha\simeq0.4$, whereas the weak aligned cut is tracked by a single slow family with large neighboring-overlap values.  The same trend appears in the $5\times5$, $6\times6$, and $7\times7$ finite-size checks summarized in Appendix~\ref{app:geometry}, but we keep the claim at the level of finite-size evidence.

Case C, $(r_x,r_y)=(4,4)$, produces corner-biased slow modes.  For the mode-profile panel we use a slightly rectangular $6\times5$ sample to make the corner bias visually distinguishable, whereas the response maps and finite-size checks use the $6\times6$ geometry.  For a $6\times5$ rectangle at $\alpha_x=\alpha_y=0$, the right gap-mode diagonal weight has center of mass $(4.02,4.01)$, while the left one has center of mass $(2.29,3.00)$.  The maximum right weight is at site $(6,5)$ and the maximum left weight is at site $(1,3)$.  The modes therefore have opposing spatial bias, but they are not ideal delta functions at opposite corners on this finite lattice.  The Petermann factor is $K\simeq7.42$.  The gap map in Fig.~\ref{fig:geometry_2d}(d) has a minimum $0.608$ at $(\alpha_x,\alpha_y)=(0,0)$ and a maximum $1.51$ at $(0.8,0.8)$.  The cooperative part $C_{\rm coop}$, defined in Appendix~\ref{app:geometry}, is positive throughout the sampled large-link region and reaches $0.881$ at the same grid point as the gap maximum, showing that the two-link response on this finite grid is not a simple sum of two independent one-link effects.  Table~\ref{tab:geometry_response} lists the open-rectangle response measures.

The clean directional boundary-control case is the cylinder that is open in $x$ and periodic in $y$ for case A.  Closing the open $x$ direction changes the gap by about $0.999$, while twisting the already periodic $y$ direction changes it by less than $5\times10^{-11}$.  The cylinder therefore separates a genuine boundary closure from a harmless twist in a direction that is already periodic.  This contrasts with the open rectangle, where the global-gap response is governed by slow-channel competition.  The full construction and the limitation of the winding diagnostic are summarized in Appendix~\ref{app:geometry}.

\section{Spectral Gap and Observed Relaxation}

The inverse gap $1/\Gap$ fixes the slowest spectral time scale, but a measured relaxation curve reveals this scale only when the corresponding gap family has appreciable visibility for the selected initial state and observable.  To test this distinction, we used the one-dimensional open chain with $N=30$, $J=1$, $\Gamma=1.7$, and $r=4$.  The initial perturbations were local density-matrix perturbations at the left edge, the bulk, and the right edge.  The measured observable was a left-right imbalance, denoted $O_{\rm LR}$.  The details of the fitting window and the finite-size validation are given in Appendix~\ref{app:relaxation}.

\begin{figure*}[t]
\includegraphics[width=0.95\textwidth]{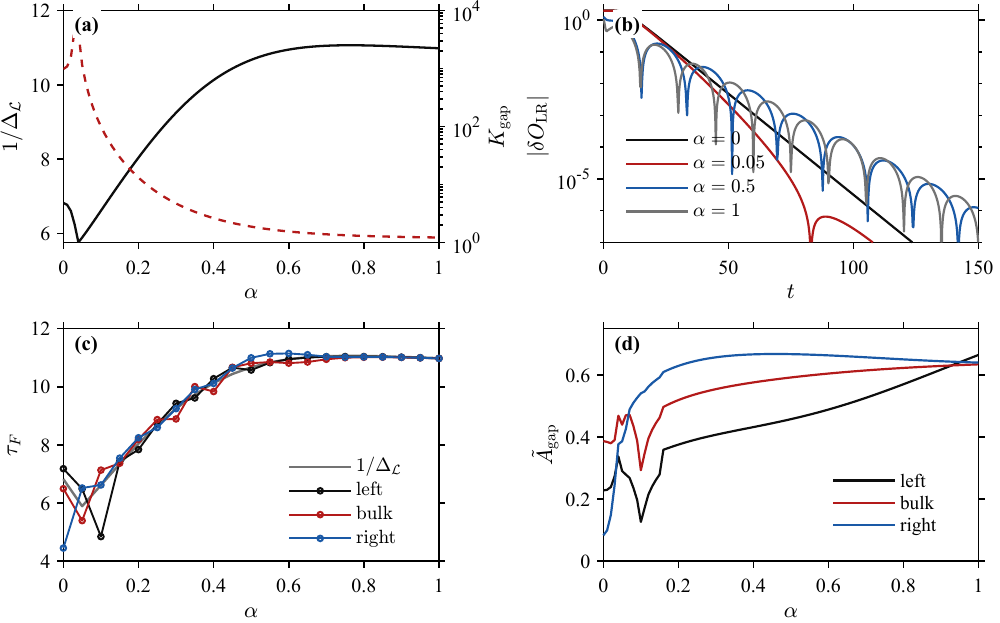}
\caption{Spectral gap versus observed relaxation for the one-dimensional open chain with $N=30$, $J=1$, $\Gamma=1.7$, and $r=4$.  (a) Inverse gap and Petermann factor of the gap mode as functions of $\alpha$.  (b) Representative time traces of $|\delta O_{\rm LR}(t)|$ for a left-edge perturbation; oscillatory zeros can affect an observable-envelope fit.  (c) Frobenius-defined late-time scale $\tau_F$ compared with $1/\Gap$ for left, bulk, and right initial perturbations.  (d) Normalized gap-family visibility $\widetilde A_{\rm gap}$ for the same perturbation-observable protocols.}
\label{fig:relaxation}
\end{figure*}

The point can be seen directly from the biorthogonal mode expansion.  For a finite diagonalizable Liouvillian, the response of an observable $O$ to an initial perturbation $\delta\rho_0$ has the form
\begin{equation}
\delta O(t)
=
\sum_{n\neq0}
e^{z_n t}
(L_n|\delta\rho_0)
\Tr(O R_n),
\label{eq:mode_response_main}
\end{equation}
where $R_n$ and $L_n$ are right and left eigenoperators.  The protocol-dependent visibility amplitude is
\begin{equation}
A_n(O,\delta\rho_0)
=
(L_n|\delta\rho_0)\Tr(O R_n).
\label{eq:mode_visibility_main}
\end{equation}
A gap mode with small $A_n$ may fail to dominate the accessible relaxation window, so a fitted relaxation time can differ from $1/\Gap$.  The normalized visibility used in Fig.~\ref{fig:relaxation}(d) and the fitting procedure are defined in Appendix~\ref{app:relaxation}.

Figure~\ref{fig:relaxation}(a) compares $1/\Gap$ with the Petermann factor of the gap mode as $\alpha$ is varied.  The gap ranges from $0.0904$ to $0.174$, so $1/\Gap$ ranges from $5.76$ to $11.1$.  The Petermann factor ranges from $1.22$ to $4.74\times10^3$.  The largest nonorthogonality occurs near, but not exactly at, the maximum of the gap.  This already shows that eigenvalue decay and eigenvector geometry are distinct pieces of information.  Representative full values are given in Table~\ref{tab:relaxation_selected}.

Figure~\ref{fig:relaxation}(b) shows representative relaxation of $|\delta O_{\rm LR}(t)|$ for a left-edge perturbation.  Oscillatory zeros in this observable can affect a direct envelope extraction, so Fig.~\ref{fig:relaxation}(c) compares $1/\Gap$ with a fitted Frobenius relaxation scale $\tau_F$ extracted from $\|\delta\rho(t)\|_F$.  The differences are protocol dependent.  At $\alpha=0$, for example, $1/\Gap\simeq6.82$, whereas the right-edge perturbation gives $\tau_F\simeq4.45$ and a normalized gap-family visibility only about $0.083$.  At $\alpha=1$, the fitted Frobenius scales for left, bulk, and right perturbations collapse close to $1/\Gap\simeq11.0$.  These values do not contradict the spectral gap.  They reflect the finite visibility of the gap family in the chosen perturbation-observable pair.

Figure~\ref{fig:relaxation}(d) makes this interpretation explicit by plotting a normalized gap-family visibility.  The visibility is not universal; it depends on both the initial perturbation and the measured observable.  Thus a boundary parameter may improve the spectral gap while not producing the same improvement in every experimentally accessible relaxation curve.

This distinction is essential for interpreting coherent boundary control in finite open systems.  A larger spectral gap means a shorter intrinsic spectral time, but a measured local relaxation curve can follow a different apparent time scale when the slow family has small projection onto the chosen perturbation-observable pair.  Figures~\ref{fig:boundary_alpha} and \ref{fig:relaxation} should therefore be read together: the self-energy controls the eigenvalue motion, while the biorthogonal mode geometry controls the visibility of that motion in a particular protocol.

\section{Conclusion}

We have analyzed a coherence-mediated mechanism by which coherent boundary conditions control finite-size Liouvillian gaps in nonreciprocal open quantum systems.  The periodic reference system admits an exact fixed-momentum reduction, from which the classical drift and the coherence-corrected diffusion follow.  In open finite systems, the coherent boundary link leaves the direct population generator unchanged but modifies the population-coherence self-energy.  Exact diagonalization and sparse near-zero eigensolvers show that this self-energy can alter the gap in one dimension and produce geometry-dependent responses in two dimensions.  Relaxation simulations further show that the inverse gap and an observed late-time scale need not coincide when the gap family has weak visibility in the chosen protocol.  This mechanism provides a way to tune slow Liouvillian modes through coherent boundary conditions rather than through a direct redesign of the dissipative jump network.

Several extensions remain natural.  A dedicated finite-size scaling analysis would be needed to determine how the open-boundary gap curves evolve toward larger systems.  It would also be useful to develop a full Liouvillian topological characterization beyond the classical population-branch winding used here as a diagnostic, and to design perturbation-observable protocols that maximize the visibility of the coherence-dressed slow modes.

\section*{Data Availability}

The data and scripts that support the findings of this work are available from the corresponding author upon reasonable request.

\begin{acknowledgments}
This work was supported by the National Natural Science Foundation of China under Grant Nos. 11975166 and 12275193.
\end{acknowledgments}

\appendix

\section{Vectorization and Numerical Liouvillian}
\label{app:vectorization}

Let the single-particle Hilbert-space dimension be $V$.  In one dimension $V=N$, and in two dimensions $V=L_xL_y$.  We use the matrix units
\begin{equation}
E_{mn}=\op{m}{n},
\qquad
1\leq m,n\leq V,
\label{eq:matrix_units_app}
\end{equation}
and expand the density matrix as
\begin{equation}
\rho=\sum_{m,n=1}^{V}\rho_{mn}E_{mn}.
\label{eq:rho_expand_app}
\end{equation}
The vector index associated with $E_{mn}$ is
\begin{equation}
\ell(m,n)=m+(n-1)V.
\label{eq:vec_index_app}
\end{equation}
Thus $\operatorname{vec}(\rho)_{\ell(m,n)}=\rho_{mn}$.

The numerical construction uses column vectorization,
\begin{equation}
\operatorname{vec}(A\rho B)
=
(B^T\otimes A)\operatorname{vec}(\rho).
\label{eq:vec_rule}
\end{equation}
The index proof is as follows.  The $(m,n)$ element of $A\rho B$ is
\begin{equation}
(A\rho B)_{mn}
=
\sum_{a,b}A_{ma}\rho_{ab}B_{bn}.
\label{eq:vec_proof_1_app}
\end{equation}
The $\ell(m,n)$ component of $(B^T\otimes A)\operatorname{vec}(\rho)$ is
\begin{equation}
\sum_{a,b}(B^T)_{nb}A_{ma}\rho_{ab}
=
\sum_{a,b}B_{bn}A_{ma}\rho_{ab},
\label{eq:vec_proof_2_app}
\end{equation}
which is Eq.~\eqref{eq:vec_proof_1_app}.  This gives Eq.~\eqref{eq:vec_rule}.

With this convention, the Hamiltonian part becomes
\begin{equation}
\operatorname{vec}[-\ii(H\rho-\rho H)]
=
-\ii
\left(
I\otimes H
-H^T\otimes I
\right)
\operatorname{vec}(\rho),
\label{eq:vec_ham}
\end{equation}
and a single jump operator $L$ contributes
\begin{equation}
(L^\dagger)^T\otimes L
-\frac{1}{2}I\otimes L^\dagger L
-\frac{1}{2}(L^\dagger L)^T\otimes I.
\label{eq:vec_jump}
\end{equation}
Equations~\eqref{eq:vec_ham} and \eqref{eq:vec_jump} define the finite Liouvillian matrix used in the calculations.

The trace functional is represented by
\begin{equation}
\langle\!\langle I|
=
\operatorname{vec}(I)^T.
\label{eq:trace_row_app}
\end{equation}
Trace preservation is the left-null-vector condition
\begin{equation}
\langle\!\langle I|\cL=0.
\label{eq:trace_preservation_app}
\end{equation}
The numerical residual is
\begin{equation}
\epsilon_{\rm tr}
=
\|\operatorname{vec}(I)^T\cL\|_2.
\label{eq:trace_residual_app}
\end{equation}
Trace preservation was checked by applying the left trace vector to the constructed matrix.  The maximum trace residuals reported in the periodic, boundary-link, geometry, and relaxation calculations are at or below the $10^{-15}$ level for the dense validations and remain within the stated sparse tolerance for larger calculations.

\section{Periodic Fixed-Momentum Reduction}
\label{app:periodic_reduction}

For a periodic lattice, $x_\mu+L_\mu$ is identified with $x_\mu$.  The allowed center-of-mass momenta are
\begin{equation}
Q_\mu=\frac{2\pi n_\mu}{L_\mu},
\qquad
n_\mu=0,1,\ldots,L_\mu-1.
\label{eq:allowed_Q_app}
\end{equation}
The relative coordinate $\bfr$ is also understood modulo the periodic lattice.  Equation~\eqref{eq:O_Qr} is a complete basis for a periodic lattice.  The orthogonality relation is
\begin{equation}
\Tr[O_{\bfQ,\bfr}^\dagger O_{\bfQ',\bfr'}]
=
V\delta_{\bfQ,\bfQ'}\delta_{\bfr,\bfr'}.
\label{eq:O_orthogonality}
\end{equation}
To see this explicitly, write $\Delta\bfQ=\bfQ'-\bfQ$ and
\begin{align*}
&\Tr[O_{\bfQ,\bfr}^\dagger O_{\bfQ',\bfr'}]
\nonumber\\
&=
\sum_{\bfx,\bfy}
e^{-\ii\bfQ\cdot\bfx}
e^{\ii\bfQ'\cdot\bfy}
\Tr[
\op{\bfx}{\bfx+\bfr}
\op{\bfy+\bfr'}{\bfy}
]
\nonumber\\
&=
\sum_{\bfx}
e^{\ii\Delta\bfQ\cdot\bfx}
\delta_{\bfx+\bfr,\bfx+\bfr'}
\nonumber\\
&=
\delta_{\bfr,\bfr'}
\sum_{\bfx}
e^{\ii\Delta\bfQ\cdot\bfx}.
\end{align*}
which gives Eq.~\eqref{eq:O_orthogonality}.

For $\bfr=0$, the gain term of a $+\mu$ jump gives
\begin{align}
\sum_{\bfx}
L_{\bfx,\mu,+}O_{\bfQ,0}L_{\bfx,\mu,+}^\dagger
&=
\gamma_{\mu,+}e^{-\ii Q_\mu}O_{\bfQ,0},
\label{eq:gain_plus}
\end{align}
where the final phase follows by relabeling $\bfx+\ehat_\mu$ as the output site.  The $-\mu$ jump gives $\gamma_{\mu,-}e^{\ii Q_\mu}O_{\bfQ,0}$.  Subtracting the loss term gives Eq.~\eqref{eq:a_Q}.  For $\bfr\neq0$, the ket and bra cannot both satisfy the same source-site condition in the gain term, so the gain vanishes and the coherence has only the loss $-\Gamma$ before the Hamiltonian is included.  In matrix-unit language, the product
\begin{equation}
\op{\bfx+s\ehat_\mu}{\bfx}
\op{\bfy+\bfr}{\bfy}
\op{\bfx}{\bfx+s\ehat_\mu}
\label{eq:coherence_gain_condition_app}
\end{equation}
is nonzero only if $\bfx=\bfy+\bfr$ and $\bfy=\bfx$ simultaneously.  These two conditions imply $\bfr=0$.

The Hamiltonian part changes the relative coordinate.  For a nearest-neighbor periodic Hamiltonian,
\begin{align}
-\ii[H,O_{\bfQ,\bfr}]
&=
-\ii\sum_\mu J_\mu
(1-e^{\ii Q_\mu})O_{\bfQ,\bfr+\ehat_\mu}
\nonumber\\
&\quad
-\ii\sum_\mu J_\mu
(1-e^{-\ii Q_\mu})O_{\bfQ,\bfr-\ehat_\mu}.
\label{eq:H_Qr_action}
\end{align}
Thus the population site $\bfr=0$ is coupled to a lattice of coherences in relative-coordinate space.

The two terms in Eq.~\eqref{eq:H_Qr_action} come from left and right multiplication.  Left multiplication gives
\begin{equation}
HO_{\bfQ,\bfr}
=
\sum_\mu J_\mu
\left(
O_{\bfQ,\bfr+\ehat_\mu}
+
O_{\bfQ,\bfr-\ehat_\mu}
\right),
\label{eq:H_left_app}
\end{equation}
whereas right multiplication gives
\begin{equation}
O_{\bfQ,\bfr}H
=
\sum_\mu J_\mu
\left(
e^{\ii Q_\mu}O_{\bfQ,\bfr+\ehat_\mu}
+
e^{-\ii Q_\mu}O_{\bfQ,\bfr-\ehat_\mu}
\right).
\label{eq:H_right_app}
\end{equation}
The phases in Eq.~\eqref{eq:H_right_app} arise because the bra coordinate is shifted and then relabeled.  Equations~\eqref{eq:H_left_app} and \eqref{eq:H_right_app} also show why a truncation to nearest-neighbor coherences is not exact: once a nearest-neighbor coherence is populated, the Hamiltonian connects it to longer-range coherences.

The relative-momentum derivation follows by defining
\begin{equation}
B_{\bfQ,\bfp}=\ket{\bfp+\bfQ}\bra{\bfp}.
\label{eq:B_Qp}
\end{equation}
The Hamiltonian is diagonal in this basis, while the jump gain is rank one in $\bfp$.  This gives
\begin{equation}
\cL B_{\bfQ,\bfp}
=
\left[-\Gamma-\ii(\varepsilon_{\bfp+\bfQ}-\varepsilon_{\bfp})\right]B_{\bfQ,\bfp}
+
\frac{c_{\bfQ}}{V}\sum_{\bfp'}B_{\bfQ,\bfp'}.
\label{eq:L_BQp}
\end{equation}
Equivalently, in the basis $B_{\bfQ,\bfp}$ the fixed-$\bfQ$ matrix has elements
\begin{equation}
(\cL_{\bfQ})_{\bfp\bfp'}
=
\left[
-\Gamma-\ii(\varepsilon_{\bfp+\bfQ}-\varepsilon_{\bfp})
\right]\delta_{\bfp\bfp'}
+
\frac{c_{\bfQ}}{V}.
\label{eq:LQ_matrix_app}
\end{equation}
Define
\begin{align}
d_{\bfp}(z,\bfQ)
&=
z+\Gamma+\ii(\varepsilon_{\bfp+\bfQ}-\varepsilon_{\bfp}),
\label{eq:d_p_Q_app}\\
A_{\bfQ}(z)
&=
\frac{1}{V}\sum_{\bfp}\frac{1}{d_{\bfp}(z,\bfQ)}.
\label{eq:d_A_Q_app}
\end{align}
Since $z-\cL_{\bfQ}$ is a diagonal matrix minus a rank-one matrix, the Sherman-Morrison formula gives
\begin{align}
[(z-\cL_{\bfQ})^{-1}]_{\bfp\bfp'}
&=
\frac{\delta_{\bfp\bfp'}}{d_{\bfp}}
\nonumber\\
&\quad+
\frac{c_{\bfQ}/V}{1-c_{\bfQ}A_{\bfQ}(z)}
\frac{1}{d_{\bfp}d_{\bfp'}}.
\label{eq:periodic_resolvent_app}
\end{align}
The pole condition $1-c_{\bfQ}A_{\bfQ}(z)=0$ is precisely the scalar secular equation in Eq.~\eqref{eq:secular}.  Writing an eigenoperator as $\sum_{\bfp}b_{\bfp}B_{\bfQ,\bfp}$ and summing over $\bfp$ gives the same condition.  Expanding it to second order in $\bfQ$ gives Eqs.~\eqref{eq:hydro} to \eqref{eq:D_eff_cosine}.

\section{Momentum Representation and Hydrodynamic Expansion}
\label{app:momentum_hydro}

We now give the small-$\bfQ$ expansion in full.  The fixed-$\bfQ$ eigenoperator is
\begin{equation}
\Psi_{\bfQ}
=
\sum_{\bfp}b_{\bfp}B_{\bfQ,\bfp}.
\label{eq:Psi_Q_app}
\end{equation}
Using Eq.~\eqref{eq:L_BQp}, the eigenvalue equation $\cL\Psi_{\bfQ}=z\Psi_{\bfQ}$ becomes
\begin{equation}
\left[
z+\Gamma+\ii(\varepsilon_{\bfp+\bfQ}-\varepsilon_{\bfp})
\right]b_{\bfp}
=
\frac{c_{\bfQ}}{V}S,
\qquad
S=\sum_{\bfp'}b_{\bfp'}.
\label{eq:bp_equation_app}
\end{equation}
For the population-connected branch, $S\neq0$.  Dividing by the bracket and summing over $\bfp$ gives
\begin{equation}
1
=
c_{\bfQ}
\left\langle
\frac{1}{
z+\Gamma+\ii(\varepsilon_{\bfp+\bfQ}-\varepsilon_{\bfp})
}
\right\rangle_{\bfp},
\label{eq:secular_app_again}
\end{equation}
where $\langle\cdots\rangle_{\bfp}$ denotes the average over the Brillouin zone.

Write
\begin{equation}
z=z_1+z_2+O(Q^3),
\qquad
z_1=O(Q),
\qquad
z_2=O(Q^2).
\label{eq:z_expand_app}
\end{equation}
The coefficient $c_{\bfQ}$ has the expansion
\begin{align}
c_{\bfQ}
&=
\Gamma
-\ii\sum_\mu(\gamma_{\mu,+}-\gamma_{\mu,-})Q_\mu
\nonumber\\
&\quad
-\frac{1}{2}
\sum_\mu(\gamma_{\mu,+}+\gamma_{\mu,-})Q_\mu^2
+O(Q^3)
\nonumber\\
&=
\Gamma+c_1+c_2+O(Q^3),
\label{eq:c_expand_app}
\end{align}
with
\begin{equation}
c_1=-\ii\sum_\mu v_\mu Q_\mu,
\qquad
c_2=-\frac{1}{2}\sum_\mu\Gamma_\mu Q_\mu^2,
\label{eq:c1_c2_app}
\end{equation}
where $v_\mu=\gamma_{\mu,+}-\gamma_{\mu,-}$ and $\Gamma_\mu=\gamma_{\mu,+}+\gamma_{\mu,-}$.

The dispersion difference is
\begin{equation}
\varepsilon_{\bfp+\bfQ}-\varepsilon_{\bfp}
=
\sum_\mu u_\mu(\bfp)Q_\mu
+
\frac{1}{2}
\sum_{\mu,\nu}M_{\mu\nu}(\bfp)Q_\mu Q_\nu
+O(Q^3),
\label{eq:delta_epsilon_expand_app}
\end{equation}
where
\begin{equation}
u_\mu(\bfp)=\partial_{p_\mu}\varepsilon_{\bfp},
\qquad
M_{\mu\nu}(\bfp)=\partial_{p_\mu}\partial_{p_\nu}\varepsilon_{\bfp}.
\label{eq:u_M_app}
\end{equation}
For a periodic band,
\begin{equation}
\langle u_\mu\rangle_{\bfp}=0,
\qquad
\langle M_{\mu\nu}\rangle_{\bfp}=0.
\label{eq:periodic_derivative_average_app}
\end{equation}
Define
\begin{equation}
I(z,\bfQ)
=
\left\langle
\frac{1}{
\Gamma+z+\ii(\varepsilon_{\bfp+\bfQ}-\varepsilon_{\bfp})
}
\right\rangle_{\bfp}.
\label{eq:I_app}
\end{equation}
Using
\begin{equation}
\frac{1}{\Gamma+A}
=
\frac{1}{\Gamma}
-\frac{A}{\Gamma^2}
+\frac{A^2}{\Gamma^3}
+O(A^3),
\label{eq:denominator_expansion_app}
\end{equation}
with $A=z+\ii(\varepsilon_{\bfp+\bfQ}-\varepsilon_{\bfp})$, we find
\begin{equation}
I
=
\frac{1}{\Gamma}
-\frac{z}{\Gamma^2}
+
\frac{
z_1^2
-
\left\langle
\left(\sum_\mu u_\mu Q_\mu\right)^2
\right\rangle_{\bfp}
}{\Gamma^3}
+O(Q^3).
\label{eq:I_expand_app}
\end{equation}
The secular equation is $1=c_{\bfQ}I$.  At first order,
\begin{equation}
0=\frac{c_1-z_1}{\Gamma},
\qquad
z_1=c_1=-\ii\sum_\mu v_\mu Q_\mu.
\label{eq:first_order_hydro_app}
\end{equation}
At second order,
\begin{equation}
0
=
\frac{c_2-z_2}{\Gamma}
-\frac{c_1z_1}{\Gamma^2}
+
\frac{
z_1^2
-
\left\langle
\left(\sum_\mu u_\mu Q_\mu\right)^2
\right\rangle_{\bfp}
}{\Gamma^2}.
\label{eq:second_order_hydro_app}
\end{equation}
Since $z_1=c_1$, the terms $-c_1z_1$ and $z_1^2$ cancel.  Therefore
\begin{equation}
z_2
=
c_2
-
\frac{1}{\Gamma}
\left\langle
\left(\sum_\mu u_\mu Q_\mu\right)^2
\right\rangle_{\bfp}.
\label{eq:z2_app}
\end{equation}
Combining Eqs.~\eqref{eq:first_order_hydro_app} and \eqref{eq:z2_app} gives Eqs.~\eqref{eq:hydro} and \eqref{eq:D_eff_general}.  For $\varepsilon_{\bfp}=2\sum_\mu J_\mu\cos p_\mu$,
\begin{equation}
u_\mu(\bfp)=-2J_\mu\sin p_\mu,
\qquad
\langle u_\mu u_\nu\rangle_{\bfp}=2J_\mu^2\delta_{\mu\nu},
\label{eq:cosine_average_app}
\end{equation}
which gives Eq.~\eqref{eq:D_eff_cosine}.

\section{Feshbach Self-Energy}
\label{app:feshbach}

The main text gives the minimal block-elimination argument behind Eqs.~\eqref{eq:block_eigen_main}--\eqref{eq:intro_self_energy}.  Here we give the matrix-unit derivation showing explicitly how the coherent boundary parameter enters the projected blocks.  In one dimension the boundary part of the Hamiltonian is
\begin{equation}
H_b(\alpha)
=
J\left(
\alpha E_{1N}+\alpha^*E_{N1}
\right).
\label{eq:H_boundary_app}
\end{equation}
Acting on the two boundary populations gives
\begin{align}
-\ii[H_b,E_{NN}]
&=
-\ii J\alpha E_{1N}
+\ii J\alpha^*E_{N1},
\label{eq:Hb_ENN_app}\\
-\ii[H_b,E_{11}]
&=
-\ii J\alpha^*E_{N1}
+\ii J\alpha E_{1N}.
\label{eq:Hb_E11_app}
\end{align}
For any bulk population $E_{xx}$ with $2\leq x\leq N-1$, the boundary Hamiltonian gives zero.  Equations~\eqref{eq:Hb_ENN_app} and \eqref{eq:Hb_E11_app} map populations to coherences, not to populations.  Hence $H_b(\alpha)$ does not modify the direct population block $\cL_{pp}$.

The reverse process is visible by acting on the two boundary coherences:
\begin{align}
-\ii[H_b,E_{N1}]
&=
-\ii J\alpha(E_{11}-E_{NN}),
\label{eq:Hb_EN1_app}\\
-\ii[H_b,E_{1N}]
&=
-\ii J\alpha^*(E_{NN}-E_{11}).
\label{eq:Hb_E1N_app}
\end{align}
These terms map coherences back into populations and therefore contribute to $\cL_{pc}$.  Other coherences are either mapped within the coherence sector or are unaffected by the boundary term.  This matrix-unit calculation is the local algebra behind Fig.~\ref{fig:mechanism}: the boundary link changes the population problem only after the coherence sector has been eliminated.

For completeness, we also spell out the finite-system block derivation used in the numerical diagnostics.  The finite open-chain calculation uses the projectors
\begin{equation}
\cP\rho=\sum_x \op{x}{x}\rho_{xx},
\qquad
\cQ=1-\cP.
\label{eq:PQ_projectors}
\end{equation}
In the ordered vector $(\cP\rho,\cQ\rho)^T$, the Liouvillian has the block form
\begin{equation}
\cL
=
\begin{pmatrix}
\cL_{pp} & \cL_{pc}\\
\cL_{cp} & \cL_{cc}
\end{pmatrix}.
\label{eq:block_L}
\end{equation}
For a right eigenvector $(p,c)^T$ with eigenvalue $z$,
\begin{align}
\cL_{pp}p+\cL_{pc}c&=zp,
\label{eq:fesh_1}\\
\cL_{cp}p+\cL_{cc}c&=zc.
\label{eq:fesh_2}
\end{align}
If $\cL_{cc}-z$ is invertible, Eq.~\eqref{eq:fesh_2} gives
\begin{equation}
c=-(\cL_{cc}-z)^{-1}\cL_{cp}p.
\label{eq:c_elim}
\end{equation}
Substituting this into Eq.~\eqref{eq:fesh_1} gives
\begin{equation}
\left[
\cL_{pp}
-
\cL_{pc}(\cL_{cc}-z)^{-1}\cL_{cp}
\right]p
=
zp.
\label{eq:fesh_eff}
\end{equation}
This is the finite-system version of Eq.~\eqref{eq:intro_self_energy}.  In the one-dimensional boundary-link diagnostic, the maximum value of $\|\cL_{pp}(\alpha)-\cL_{pp}(0)\|_F$ over the scan was zero within machine precision.  The plotted self-energy diagnostic uses the self-consistent gap eigenvalue $z_{\rm gap}(\alpha)$ at each point and compares
\begin{equation}
\delta\Sigma_{\rm sc}(\alpha)
=
\Sigma[z_{\rm gap}(\alpha),\alpha]
-
\Sigma[z_{\rm gap}(0),0].
\label{eq:delta_sigma_sc_app}
\end{equation}
The numerical implementation evaluates the equivalent resolvent-return matrix with $(z-\cL_{cc})^{-1}$ rather than $(\cL_{cc}-z)^{-1}$; this changes the overall sign of $\Sigma$ but not the Frobenius norm of $\delta\Sigma_{\rm sc}$.  With this convention the self-energy difference changed by as much as $2.543519$ for $N=12$, $r=4$, and $\Gamma=1.7$.  The solve residual in the Feshbach construction was $1.010347\times10^{-15}$.

The sign convention is worth spelling out.  Starting from Eq.~\eqref{eq:fesh_2}, one has
\begin{equation}
(\cL_{cc}-z)c=-\cL_{cp}p.
\label{eq:fesh_sign_app}
\end{equation}
Thus
\begin{equation}
c=-(\cL_{cc}-z)^{-1}\cL_{cp}p.
\label{eq:fesh_c_solution_app}
\end{equation}
Substitution into Eq.~\eqref{eq:fesh_1} gives
\begin{equation}
\cL_{pp}p
-
\cL_{pc}(\cL_{cc}-z)^{-1}\cL_{cp}p
=
zp.
\label{eq:fesh_sign_final_app}
\end{equation}
This is why Eq.~\eqref{eq:intro_self_energy} appears with a minus sign in the effective population Liouvillian.  Because the full Liouvillian is not Hermitian, the two off-diagonal blocks $\cL_{pc}$ and $\cL_{cp}$ should be kept as distinct matrices.  Writing the self-energy as $V^\dagger(\cL_{cc}-z)^{-1}V$ would assume an additional Hermitian relation that is not generally present.

\subsection{Weak-link estimate for the selected slow branch}

The same boundary-control mechanism can be phrased as a finite-size perturbation problem near the open coherent boundary, $\alpha=0$.  In the one-dimensional scan the dissipative graph is fixed and only the coherent boundary Hamiltonian is varied.  For real $\alpha$,
\begin{equation}
\cL(\alpha)
=
\cL_0+\alpha\mathcal V_{\rm bd},
\qquad
\cL_0=\cL(0),
\label{eq:L_weak_link_app}
\end{equation}
where
\begin{equation}
\mathcal V_{\rm bd}\rho
=
-\ii J[h_{\rm bd},\rho],
\qquad
h_{\rm bd}
=
\op{1}{N}+\op{N}{1}.
\label{eq:Vbd_action_app}
\end{equation}
This is a perturbation of the full finite Liouvillian, not of the population block alone.  The latter remains independent of $\alpha$, as shown above.

Let $z_g^{(0)}$ be a nonzero isolated eigenvalue of $\cL_0$ that is connected to the branch controlling the gap at small $\alpha$.  The right and left eigenvectors are defined by
\begin{align}
\cL_0|R_n^{(0)})
&=
z_n^{(0)}|R_n^{(0)}),
\label{eq:right_weak_app}\\
(L_n^{(0)}|\cL_0
&=
z_n^{(0)}(L_n^{(0)}|,
\label{eq:left_weak_app}
\end{align}
with biorthogonal normalization
\begin{equation}
(L_m^{(0)}|R_n^{(0)})
=
\delta_{mn}.
\label{eq:biorth_weak_app}
\end{equation}
For a diagonalizable finite matrix and an isolated branch, the branch eigenvalue has the local expansion
\begin{equation}
z_g(\alpha)
=
z_g^{(0)}
+
\alpha z_g^{(1)}
+
\alpha^2 z_g^{(2)}
+
O(\alpha^3).
\label{eq:z_expansion_weak_app}
\end{equation}
The first two coefficients are
\begin{equation}
z_g^{(1)}
=
(L_g^{(0)}|\mathcal V_{\rm bd}|R_g^{(0)}),
\label{eq:z1_weak_app}
\end{equation}
and
\begin{equation}
z_g^{(2)}
=
\sum_{m\neq g}
\frac{
(L_g^{(0)}|\mathcal V_{\rm bd}|R_m^{(0)})
(L_m^{(0)}|\mathcal V_{\rm bd}|R_g^{(0)})
}{
z_g^{(0)}-z_m^{(0)}
}.
\label{eq:z2_weak_app}
\end{equation}
Equation~\eqref{eq:z2_weak_app} is meaningful only when the branch is sufficiently isolated from the rest of the spectrum and the left-right basis is not too ill conditioned.

For the branch decay rate,
\begin{equation}
\Delta_g(\alpha)
=
-\operatorname{Re}z_g(\alpha),
\label{eq:Delta_branch_weak_app}
\end{equation}
one obtains
\begin{equation}
\Delta_g(\alpha)
=
\Delta_g(0)
+
a_g\alpha
+
b_g\alpha^2
+
O(\alpha^3),
\label{eq:Delta_expansion_weak_app}
\end{equation}
with
\begin{equation}
a_g=-\operatorname{Re}z_g^{(1)},
\qquad
b_g=-\operatorname{Re}z_g^{(2)}.
\label{eq:ab_weak_app}
\end{equation}
If $0<\alpha_{\max}^{(g)}<1$ and the same branch remains the one selected by the lower envelope, the local single-branch estimate is
\begin{equation}
\alpha_{\max}^{(g)}
\simeq
-
\frac{\operatorname{Re}z_g^{(1)}}
{2\operatorname{Re}z_g^{(2)}}.
\label{eq:alpha_max_weak_app}
\end{equation}
This formula is a finite-size weak-link estimate.  It is not an exact law for the full scan, and it should not be used when the maximum lies outside the small-$\alpha$ regime or when the relevant slow family changes before the stationary point is reached.

The lower-envelope form of the gap also allows a local two-branch interpretation.  Suppose two competing finite-size branches have
\begin{align}
\Delta_a(\alpha)
&=
\Delta_a^{(0)}
+
a_a\alpha
+
b_a\alpha^2,
\label{eq:two_branch_a_app}\\
\Delta_b(\alpha)
&=
\Delta_b^{(0)}
+
a_b\alpha
+
b_b\alpha^2.
\label{eq:two_branch_b_app}
\end{align}
Their local crossing is determined by
\begin{equation}
(b_a-b_b)\alpha_c^2
+
(a_a-a_b)\alpha_c
+
(\Delta_a^{(0)}-\Delta_b^{(0)})
=
0.
\label{eq:two_branch_crossing_app}
\end{equation}
A maximum in the lower envelope can occur near such a crossing when the branch selected before the crossing has positive slope and the branch selected after the crossing has negative slope.  This is again an interpretive finite-size estimate, not a theorem.

Table~\ref{tab:weak_link_estimate} gives the branch-resolved weak-link diagnostics for the Fig.~\ref{fig:boundary_alpha}(a) parameters.  The entries are obtained for the finite $N=30$, $J=1$, $\Gamma=1.7$ calculation, with $\gamma_+=\Gamma r/(1+r)$ and $\gamma_-=\Gamma/(1+r)$.  The first-order coefficient is numerically negligible for all four ratios.  The second-order coefficient gives a positive curvature of the decay rate, and this curvature grows rapidly with nonreciprocity.  This explains the initial small-$\alpha$ enhancement, especially for $r=4$ and $r=8$.  However, the stationary-point condition in Eq.~\eqref{eq:alpha_max_weak_app} is not satisfied in a controlled way for these data, so the perturbative peak position is reported as not applicable.  A recursive Taylor check through eighth order gave no stable local maximum in $(0,1)$ for $r=4$ or $r=8$.  Euclidean-overlap tracking of the small-$\alpha$ gap family gave a minimum overlap above $0.999$ for these two ratios, so the observed peak is best described as a finite-size turnover of the same slow family beyond the local weak-link expansion, rather than as a resolved two-branch crossing at the scan resolution.

\begin{table*}[t]
\caption{Finite-size weak-link diagnostics for the one-dimensional boundary-link scan in Fig.~\ref{fig:boundary_alpha}(a).  Here $\alpha_{\max}^{\rm ED}$ is read from the finite-size gap curve, while $\alpha_{\max}^{\rm pert}$ denotes the single-branch weak-link stationary point in Eq.~\eqref{eq:alpha_max_weak_app} when its validity conditions are met.  N/A means that the perturbative stationary point is outside its controlled regime or is absent.}
\label{tab:weak_link_estimate}
\begin{ruledtabular}
\begin{tabular}{cccccccl}
$r$ &
$\alpha_{\max}^{\rm ED}$ &
$\alpha_{\max}^{\rm pert}$ &
$\Gap(0)$ &
$\Gap^{\max}$ &
$\operatorname{Re}z_g^{(1)}$ &
$\operatorname{Re}z_g^{(2)}$ &
interpretation \\
$1$ & $0.890$ & N/A & $0.0223366$ & $0.0906274$ & $5.93\times10^{-17}$ & $-0.314951$ & outside weak-link regime \\
$2$ & $0.110$ & N/A & $0.0618053$ & $0.0922334$ & $-1.24\times10^{-16}$ & $-1.45960$ & no controlled stationary point \\
$4$ & $0.040$ & N/A & $0.146562$ & $0.173582$ & $-3.89\times10^{-12}$ & $-10.6917$ & initial curvature only \\
$8$ & $0.020$ & N/A & $0.224688$ & $0.253709$ & $-3.52\times10^{-10}$ & $-37.5249$ & initial curvature only \\
\end{tabular}
\end{ruledtabular}
\end{table*}

\subsection{Boundary perturbation as a low-rank resolvent problem}

The connection between the periodic reference problem and a finite boundary modification can also be written as a resolvent identity.  Let
\begin{equation}
\cL(\alpha)
=
\cL_{\rm ref}
+
\delta\cL_{\rm bd}(\alpha),
\label{eq:L_low_rank_start}
\end{equation}
where $\cL_{\rm ref}$ is a reference Liouvillian and $\delta\cL_{\rm bd}$ has finite support near the boundary in Liouville space.  Any finite-rank boundary perturbation can be factored as
\begin{equation}
\delta\cL_{\rm bd}
=
U_{\rm bd}V_{\rm bd}^{\dagger}.
\label{eq:UV_boundary_app}
\end{equation}
Here $V_{\rm bd}^{\dagger}$ denotes a dual boundary map.  It is not assumed to be the Hermitian conjugate of $U_{\rm bd}$, because the Liouvillian is non-Hermitian.  Starting from
\begin{equation}
(\cL_{\rm ref}+U_{\rm bd}V_{\rm bd}^{\dagger})|\Psi)
=
z|\Psi),
\label{eq:low_rank_eig_app}
\end{equation}
one obtains
\begin{equation}
|\Psi)
=
(z-\cL_{\rm ref})^{-1}U_{\rm bd}V_{\rm bd}^{\dagger}|\Psi).
\label{eq:low_rank_resolvent_state_app}
\end{equation}
Multiplying by $V_{\rm bd}^{\dagger}$ gives a finite-dimensional equation in the boundary subspace.  Nontrivial solutions require
\begin{equation}
\det\left[
I
-
V_{\rm bd}^{\dagger}
(z-\cL_{\rm ref})^{-1}
U_{\rm bd}
\right]
=0.
\label{eq:low_rank_secular_app}
\end{equation}
This expression is the boundary analogue of the rank-one denominator in Eq.~\eqref{eq:periodic_resolvent_app}.

If $\cL_{\rm ref}$ is chosen as the fully periodic Liouvillian, removing the boundary jumps is another finite-support boundary perturbation.  In the open-boundary scans of Sec.~IV, however, the dissipative network is kept fixed, so the $\alpha$-dependent part is only the coherent Hamiltonian boundary term.  For the one-dimensional coherent boundary Hamiltonian, choose the closed coherent chain with the same dissipator as the reference and write $\theta=\alpha-1$.  The boundary perturbation is
\begin{equation}
\delta H_{\rm bd}
=
\theta J(E_{1N}+E_{N1}).
\label{eq:delta_H_boundary_app}
\end{equation}
Using Liouville-space matrix units $|mn)\equiv E_{mn}$,
\begin{align}
\delta\cL_{\rm bd}^{(H)}|mn)
&=
-\ii\theta J\delta_{m,N}|1n)
-\ii\theta J\delta_{m,1}|Nn)
\nonumber\\
&\quad
+\ii\theta J\delta_{n,1}|mN)
+\ii\theta J\delta_{n,N}|m1).
\label{eq:delta_L_boundary_action_app}
\end{align}
Therefore
\begin{align}
\delta\cL_{\rm bd}^{(H)}
&=
-\ii\theta J\sum_n |1n)(Nn|
-\ii\theta J\sum_n |Nn)(1n|
\nonumber\\
&\quad
+\ii\theta J\sum_m |mN)(m1|
+\ii\theta J\sum_m |m1)(mN|.
\label{eq:delta_L_low_rank_sum_app}
\end{align}
The rank is at most $4N$ in the $N^2$-dimensional Liouville space.  This finite support is why a coherent boundary link can be treated as a low-rank resolvent problem, even though its spectral consequences for a non-normal Liouvillian can be large.

\section{Numerical Protocol}
\label{app:numerics}

The gap was computed from the finite Liouvillian matrix using dense diagonalization for smaller systems and a sparse near-zero eigensolver for larger scans.  The zero eigenvalue was identified by a tolerance $10^{-10}$ unless stated otherwise.  The nonzero eigenvalue with the largest real part determined $\Gap$ through Eq.~\eqref{eq:gap_definition}.  Selected sparse results were checked against dense diagonalization.

For the one-dimensional boundary-link scan, dense-sparse gap validation over selected $N$, $r$, and $\alpha$ values gave a maximum difference $1.401571\times10^{-9}$.  For the relaxation analysis, the corresponding maximum dense-sparse difference was $2.207530\times10^{-10}$.  The spectral time-evolution validation compared direct matrix exponentiation with the spectral reconstruction and gave maximum relative error $1.124511\times10^{-13}$.

For a computed right eigenvector $R_n$ and left eigenvector $L_n$, the residuals are
\begin{align}
\epsilon_R
&=
\frac{\|\cL R_n-\lambda_n R_n\|_2}{\|R_n\|_2},
\label{eq:right_residual_app}\\
\epsilon_L
&=
\frac{\|\cL^\dagger L_n-\lambda_n^*L_n\|_2}{\|L_n\|_2}.
\label{eq:left_residual_app}
\end{align}
The numerical validation records dense-sparse gap comparisons, trace residuals, zero-mode counts, and the spectral time-evolution reconstruction error.  Since left-right eigenvector residuals were not evaluated for every scan point, we report the definitions here rather than quoting a single global residual maximum.

More explicitly, the finite-size gap extraction used
\begin{equation}
\Gap
=
-\max_{\lambda_n:\ |\lambda_n|>\epsilon}
\operatorname{Re}\lambda_n,
\qquad
\epsilon=10^{-10}.
\label{eq:gap_numeric_app}
\end{equation}
If several eigenvalues had the same real part within tolerance, they were treated as a gap family.  In the representative relaxation scan with $N=30$, $J=1$, $\Gamma=1.7$, and $r=4$, only 4 of the 101 sampled $\alpha$ values had a real gap mode.  The other 97 points had a complex-conjugate gap pair.  This is why the relaxation analysis keeps the whole gap family rather than only one arbitrarily selected eigenvalue.

\begin{table*}[t]
\caption{One-dimensional boundary-link scan for $N=30$, $J=1$, and $\Gamma=1.7$.  The rate ratio is $r=\gamma_+/\gamma_-$, with $\gamma_+=\Gamma r/(1+r)$ and $\gamma_-=\Gamma/(1+r)$.  The peak is the maximum over $\alpha=0,0.01,\ldots,1$.}
\label{tab:boundary_gaps}
\begin{ruledtabular}
\begin{tabular}{ccccc}
$r$ & $\Gap(0)$ & $\Gap(0.5)$ & $\Gap(1)$ & peak $(\alpha,\Gap)$\\
\hline
$1$ & $0.022336631087$ & $0.068682122137$ & $0.090586726562$ & $(0.890,0.090627396193)$\\
$2$ & $0.061805283976$ & $0.084768540003$ & $0.091507365986$ & $(0.110,0.092233433224)$\\
$4$ & $0.146561650905$ & $0.093697658991$ & $0.091034689157$ & $(0.040,0.173582162684)$\\
$8$ & $0.224687799059$ & $0.101666805657$ & $0.090766636387$ & $(0.020,0.253709058741)$
\end{tabular}
\end{ruledtabular}
\end{table*}

\begin{table*}[t]
\caption{Selected spectral and visibility data for the representative one-dimensional chain with $N=30$, $J=1$, $\Gamma=1.7$, and $r=4$.  Here $K$ is the Petermann factor of the selected gap mode, ${\rm COM}_R$ and ${\rm COM}_L$ are the one-dimensional centers of mass of the right and left diagonal weights, and $g$ is the number of modes in the gap family within tolerance.}
\label{tab:relaxation_selected}
\begin{ruledtabular}
\begin{tabular}{cccccc}
$\alpha$ & $\Gap$ & $1/\Gap$ & $K$ & $({\rm COM}_R,{\rm COM}_L)$ & $g$\\
\hline
$0$ & $0.146561650914$ & $6.823067246860$ & $9.805409\times10^2$ & $(24.307182,5.397788)$ & $1$\\
$0.05$ & $0.169742135801$ & $5.891289132660$ & $1.145601\times10^3$ & $(24.400189,5.396095)$ & $2$\\
$0.5$ & $0.093697659211$ & $10.672625212000$ & $1.871671$ & $(18.121935,13.798590)$ & $2$\\
$1$ & $0.091034689141$ & $10.984823581400$ & $1.215594$ & $(16.219572,15.295235)$ & $2$
\end{tabular}
\end{ruledtabular}
\end{table*}

\section{Geometry and Winding Diagnostic}
\label{app:geometry}

The two-dimensional calculations use
\begin{equation}
\gamma_{\mu,+}
=
\frac{\Gamma_\mu r_\mu}{1+r_\mu},
\qquad
\gamma_{\mu,-}
=
\frac{\Gamma_\mu}{1+r_\mu},
\qquad
\mu=x,y.
\label{eq:2d_rates_app}
\end{equation}
In the two-dimensional geometry calculations, $J_x=J_y=1$ and $\Gamma_x=\Gamma_y=1.7$.  For the open-rectangle gap function $\Gap(\alpha_x,\alpha_y)$, the one-direction response measures are
\begin{align}
R_x&=\max_{\alpha_x\in[0,1]}
|\Gap(\alpha_x,0)-\Gap(0,0)|,
\label{eq:Rx_app}\\
R_y&=\max_{\alpha_y\in[0,1]}
|\Gap(0,\alpha_y)-\Gap(0,0)|.
\label{eq:Ry_app}
\end{align}
The initial finite-difference responses are
\begin{align}
\chi_x
&=
\frac{\Gap(\delta\alpha,0)-\Gap(0,0)}{\delta\alpha},
\label{eq:chi_x_app}\\
\chi_y
&=
\frac{\Gap(0,\delta\alpha)-\Gap(0,0)}{\delta\alpha},
\label{eq:chi_y_app}
\end{align}
with $\delta\alpha=0.1$ in the plotted grid.

\begin{table*}[t]
\caption{Two-dimensional open-rectangle response data for the $6\times6$ scans.}
\label{tab:geometry_response}
\begin{ruledtabular}
\begin{tabular}{ccccc}
case & $(r_x,r_y)$ & $R_x$ & $R_y$ & $(\chi_x,\chi_y)$\\
\hline
A & $(4,1)$ & $0.024441048040$ & $0.142733926076$ & $(-0.007782219544,\ 0.115237089462)$\\
B & $(1,4)$ & $0.142733926063$ & $0.024441048034$ & $(0.115237089757,\ -0.007782219619)$\\
C & $(4,4)$ & $0.017225783803$ & $0.017225783788$ & $(0.054854798,\ 0.054854798)$
\end{tabular}
\end{ruledtabular}
\end{table*}

\begin{table*}[t]
\caption{Finite-size checks for the two-dimensional open-rectangle calculations.  Case A has $(r_x,r_y)=(4,1)$ and the listed values are $\Gap(\alpha_x,0)$ and $\Gap(0,\alpha_y)$ at $\alpha=0,0.5,1$.  Case C has $(r_x,r_y)=(4,4)$ and the listed values are $\Gap(\alpha,\alpha)$ at the same three $\alpha$ values.}
\label{tab:geometry_finite_size}
\begin{ruledtabular}
\begin{tabular}{cccc}
$L_x\times L_y$ & case A, $x$ cut & case A, $y$ cut & case C, diagonal\\
\hline
$5\times5$ & $(0.723084,\ 0.701772,\ 0.679996)$ & $(0.723084,\ 0.855857,\ 0.856030)$ & $(0.871804,\ 1.287368,\ 2.113947)$\\
$6\times6$ & $(0.467512,\ 0.455000,\ 0.443071)$ & $(0.467512,\ 0.610140,\ 0.610246)$ & $(0.607649,\ 0.958676,\ 1.452014)$\\
$7\times7$ & $(0.333703,\ 0.324357,\ 0.316096)$ & $(0.333703,\ 0.483566,\ 0.483603)$ & $(0.476591,\ 0.839267,\ 1.183705)$
\end{tabular}
\end{ruledtabular}
\end{table*}

Table~\ref{tab:geometry_finite_size} shows the same qualitative pattern as the main $6\times6$ figure.  For case A, varying the $y$ coherent link produces a larger gap response than varying the $x$ link at all three sizes.  For case C, increasing both coherent links enhances the gap along the diagonal cut $\alpha_x=\alpha_y$.  These checks support robustness at the finite sizes studied, but they are not a substitute for a thermodynamic finite-size scaling analysis.

\subsection{Slow-channel competition in the open rectangle}

To diagnose the lower-envelope interpretation in the open rectangle, we computed the lowest five nonzero decay rates on the same $6\times6$ cuts used in Figs.~\ref{fig:geometry_2d}(a) and \ref{fig:geometry_2d}(b).  Table~\ref{tab:slow_channels_2d} lists the lowest three rates at $\alpha=0,0.5,1$ for the responsive transverse cut and the weak aligned cut in cases A and B.  For case A, the weak aligned cut $(\alpha_x,0)$ remains on a smoothly tracked slow family: the neighboring Euclidean overlap of the selected gap vector is always larger than $0.997$.  By contrast, the responsive transverse cut $(0,\alpha_y)$ shows a slow-sector rearrangement near $\alpha_y\simeq0.4$.  The gap rises from $0.572477$ at $\alpha_y=0.3$ to $0.610130$ at $\alpha_y=0.4$, while the next decay rate is $0.658585$ at the same point.  The neighboring overlap of the selected gap vector drops to the numerical zero level at this step.  Case B shows the same behavior after interchanging $x$ and $y$.

This tracking uses ordinary Euclidean overlap between sparse-eigensolver right eigenvectors, not a biorthogonal invariant.  It is therefore used only as a diagnostic of finite-size slow-sector rearrangement.  The relevant conclusion is not a topological statement and not a thermodynamic scaling claim.  The open-rectangle transverse response is better described as lower-envelope competition among the lowest few finite-size decay channels than as a local response tied to one microscopic hopping direction.  In the data used here the selected gap eigenvalue is real on the listed cut points, so complex-conjugate gap-pair ambiguity does not affect Table~\ref{tab:slow_channels_2d}.

\begin{table*}[t]
\caption{Lowest three nonzero decay rates on the $6\times6$ open-rectangle cuts for cases A and B.  The responsive cut is transverse to the direction with nonreciprocal dissipative hopping in these finite rectangles.}
\label{tab:slow_channels_2d}
\begin{ruledtabular}
\begin{tabular}{ccccccc}
case & $(r_x,r_y)$ & cut & type & $\alpha$ & $\Delta_1$ & $(\Delta_2,\Delta_3)$\\
\hline
A & $(4,1)$ & $(\alpha_x,0)$ & weak & $0$ & $0.467512$ & $(0.610123,\ 1.060733)$\\
A & $(4,1)$ & $(\alpha_x,0)$ & weak & $0.5$ & $0.455000$ & $(0.982569,\ 1.495468)$\\
A & $(4,1)$ & $(\alpha_x,0)$ & weak & $1$ & $0.443071$ & $(1.486450,\ 1.486450)$\\
A & $(4,1)$ & $(0,\alpha_y)$ & responsive & $0$ & $0.467512$ & $(0.610123,\ 1.060733)$\\
A & $(4,1)$ & $(0,\alpha_y)$ & responsive & $0.5$ & $0.610140$ & $(0.777399,\ 1.336487)$\\
A & $(4,1)$ & $(0,\alpha_y)$ & responsive & $1$ & $0.610246$ & $(1.636341,\ 1.636341)$\\
B & $(1,4)$ & $(\alpha_x,0)$ & responsive & $0$ & $0.467512$ & $(0.610123,\ 1.060733)$\\
B & $(1,4)$ & $(\alpha_x,0)$ & responsive & $0.5$ & $0.610140$ & $(0.777399,\ 1.336487)$\\
B & $(1,4)$ & $(\alpha_x,0)$ & responsive & $1$ & $0.610246$ & $(1.636341,\ 1.636341)$\\
B & $(1,4)$ & $(0,\alpha_y)$ & weak & $0$ & $0.467512$ & $(0.610123,\ 1.060733)$\\
B & $(1,4)$ & $(0,\alpha_y)$ & weak & $0.5$ & $0.455000$ & $(0.982569,\ 1.495468)$\\
B & $(1,4)$ & $(0,\alpha_y)$ & weak & $1$ & $0.443071$ & $(1.486450,\ 1.486450)$
\end{tabular}
\end{ruledtabular}
\end{table*}

\subsection{Cooperative part of the two-link response}

For the corner-biased case C, we quantified the part of the two-link response that is not obtained by adding the two one-link cuts.  On the sampled $11\times11$ grid, define
\begin{align}
C_{\rm coop}(\alpha_x,\alpha_y)
&=
\Gap(\alpha_x,\alpha_y)
-
\Gap(\alpha_x,0)
\nonumber\\
&\quad
-
\Gap(0,\alpha_y)
+
\Gap(0,0).
\label{eq:C_coop_app}
\end{align}
This is a finite-grid diagnostic for the $6\times6$ calculation.  It is positive throughout the sampled large-link region $\alpha_x\geq0.5$ and $\alpha_y\geq0.5$, with a minimum value $0.320403$ in that region.  The maximum value is $0.881162$ at $(\alpha_x,\alpha_y)=(0.8,0.8)$, the same grid point at which $\Gap$ reaches its maximum.  At $(1,1)$ the cooperative part remains positive, $C_{\rm coop}=0.829297$.  Thus the two-link response in Fig.~\ref{fig:geometry_2d}(d) is not the sum of two independent one-link responses on this finite grid.  The corresponding summary values are listed in Table~\ref{tab:coop_2d}.

\begin{table*}[t]
\caption{Cooperative two-link diagnostic for case C, $(r_x,r_y)=(4,4)$, on the sampled $6\times6$ open-rectangle grid.}
\label{tab:coop_2d}
\begin{ruledtabular}
\begin{tabular}{ccccc}
$\Gap(0,0)$ & $\Gap^{\max}$ & $(\alpha_x,\alpha_y)_{\Gap^{\max}}$ & $\max C_{\rm coop}$ & $(\alpha_x,\alpha_y)_{C_{\rm coop}^{\max}}$\\
\hline
$0.607649$ & $1.508749$ & $(0.8,0.8)$ & $0.881162$ & $(0.8,0.8)$
\end{tabular}
\end{ruledtabular}
\end{table*}

For the two-dimensional classical population branch on a torus, the Bloch eigenvalue is
\begin{align}
a(k_x,k_y)
=&
\gamma_{x,+}(e^{-\ii k_x}-1)
+\gamma_{x,-}(e^{\ii k_x}-1)
\nonumber\\
&+
\gamma_{y,+}(e^{-\ii k_y}-1)
+\gamma_{y,-}(e^{\ii k_y}-1).
\label{eq:classical_2d_branch}
\end{align}
Holding one momentum fixed and winding the other around the Brillouin zone gives a point-gap winding diagnostic for this classical branch.  The $x$-direction diagnostic is
\begin{equation}
W_x(E_0)
=
\frac{1}{2\pi\ii}
\int_0^{2\pi}
\dd k_x\,
\partial_{k_x}\ln[a(k_x,0)-E_0],
\label{eq:Wx_app}
\end{equation}
and $W_y$ is defined analogously.  Numerically the integral is computed as the unwrapped phase change of $a-E_0$ over 400 momentum intervals.  When the corresponding ratio is unity, the diagnostic uses $E_0=-\Gamma+0.5\ii$; otherwise it uses $E_0=-\Gamma$.  In the two-dimensional geometry calculations, the diagnostics were $(W_x,W_y)=(-1,0)$ for case A, $(0,-1)$ for case B, and $(-1,-1)$ for case C.  These numbers are used only as a guide to the direction of population-branch nonreciprocity.  They are not used as a claim of a complete Liouvillian topological invariant.

The cylinder check used the same $6\times6$ size for case A, with the system open in $x$ and periodic in $y$.  Closing the open $x$ direction changed the gap by $0.998690053493$, while twisting the already periodic $y$ direction changed it by only $4.461920\times10^{-11}$.  This check separates a true boundary closure from a harmless twist in a direction that is already periodic in the finite geometry.

\section{Left-Right Modes and Relaxation Fits}
\label{app:relaxation}

The main text uses the compact response expansion in Eqs.~\eqref{eq:mode_response_main} and \eqref{eq:mode_visibility_main}.  Here we give the numerical normalization, the concrete observable, and the fitting procedure.  Let $R_n$ and $L_n$ be right and left eigenvectors of the vectorized Liouvillian, normalized so that
\begin{equation}
L_m^\dagger R_n=\delta_{mn}.
\label{eq:biorth_norm}
\end{equation}
The Petermann-type factor used in the figures is
\begin{equation}
K_n=(R_n^\dagger R_n)(L_n^\dagger L_n).
\label{eq:petermann}
\end{equation}
For a right eigenoperator $R_n$ the normalized diagonal weight is
\begin{equation}
w_R(x)=
\frac{|(R_n)_{xx}|}{\sum_y |(R_n)_{yy}|},
\label{eq:right_weight}
\end{equation}
and similarly for the left eigenoperator.  The center of mass is $\sum_x xw_{R,L}(x)$ in one dimension and the corresponding coordinate-wise sum in two dimensions.

For relaxation, the density matrix was written as
\begin{equation}
\rho(t)=\rho_{\rm ss}+\delta\rho(t),
\label{eq:rho_relax}
\end{equation}
where $\rho_{\rm ss}$ is the steady state.  The observable response was
\begin{equation}
\delta O_{\rm LR}(t)=\Tr[O_{\rm LR}\delta\rho(t)].
\label{eq:delta_O}
\end{equation}
In the relaxation analysis, the initial perturbation at site $x_0$ is
\begin{equation}
\delta\rho_{x_0}(0)
=
\op{x_0}{x_0}-\rho_{\rm ss}.
\label{eq:initial_perturbation_app}
\end{equation}
The tested sites were $x_0=1$, $x_0=\lfloor N/2\rfloor$, and $x_0=N$.  The imbalance observable is
\begin{equation}
O_{\rm LR}
=
\sum_{x=1}^{N}s_x\op{x}{x},
\qquad
s_x=
\begin{cases}
-1,&x\leq \lfloor N/2\rfloor,\\
+1,&x>\lfloor N/2\rfloor.
\end{cases}
\label{eq:imbalance_app}
\end{equation}
Using the biorthogonal eigenbasis,
\begin{equation}
|\delta\rho_{x_0}(0)\rangle\!\rangle
=
\sum_n a_n(x_0)R_n,
\qquad
a_n(x_0)=L_n^\dagger|\delta\rho_{x_0}(0)\rangle\!\rangle.
\label{eq:mode_expansion_app}
\end{equation}
The zero-mode coefficient is removed because the perturbation is measured relative to the steady state.  The observable response is then
\begin{align}
\delta O_{\rm LR}^{(x_0)}(t)
&=
\sum_{n\neq0}
\langle\!\langle O_{\rm LR}|R_n\rangle\!\rangle
a_n(x_0)e^{\lambda_n t}
\nonumber\\
&=
\sum_{n\neq0}A_n(x_0)e^{\lambda_n t},
\label{eq:observable_expansion_app}
\end{align}
where
\begin{equation}
A_n(x_0)
=
\langle\!\langle O_{\rm LR}|R_n\rangle\!\rangle
L_n^\dagger|\delta\rho_{x_0}(0)\rangle\!\rangle.
\label{eq:mode_amplitude_app}
\end{equation}
The normalized gap-family visibility plotted in Fig.~\ref{fig:relaxation}(d) is
\begin{equation}
\widetilde A_{\rm gap}(x_0)
=
\frac{
\sum_{n\in{\rm gap\ family}}|A_n(x_0)|
}{
\sum_{n\in{\rm retained\ slow\ modes}}|A_n(x_0)|
}.
\label{eq:visibility_app}
\end{equation}
The retained set contained the ten slowest modes in the visibility diagnostic.

The reported observed time $\tau_{\rm obs}$ was obtained from an exponential fit to the late-time decay window used in the relaxation analysis.  The fitting routine first works with the absolute signal $|\delta O_{\rm LR}(t)|$.  If enough local maxima are present, these maxima define the envelope.  Otherwise the absolute signal itself is fitted.  Candidate points are selected from the relative-amplitude window
\begin{equation}
10^{-3}
\leq
\frac{|\delta O_{\rm LR}(t)|}{\max_t|\delta O_{\rm LR}(t)|}
\leq
10^{-1}.
\label{eq:fit_window_app}
\end{equation}
If fewer than five points remain, the fit falls back to the last $30\%$ of the available positive data.  A linear fit of $\ln|\delta O_{\rm LR}(t)|$ or of its envelope gives a decay rate $\kappa$, and
\begin{equation}
\tau_{\rm obs}=\kappa^{-1}.
\label{eq:tau_obs_app}
\end{equation}
This fit is a protocol-dependent diagnostic and is not an alternative definition of the Liouvillian gap.

%

\end{document}